\documentclass[journal=jctcce,manuscript=article]{achemso}
\setkeys{acs}{usetitle = true} 
\usepackage[version=3]{mhchem} 
\usepackage[T1]{fontenc}       
\usepackage{amsmath,amssymb,latexsym,epsfig,multirow,tabularx}
\usepackage{amsthm}
\usepackage{verbatim}
\usepackage{siunitx} 
\usepackage{graphicx}
\usepackage[tableposition=t]{caption} 
\usepackage{subcaption}
\usepackage[titletoc,toc,title]{appendix}
\usepackage{threeparttable} 
\usepackage{wrapfig} 
\usepackage{makecell,adjustbox}
\usepackage[normalem]{ulem} 
\usepackage{xcolor,colortbl}

\author{Duc D. Nguyen}
\author{Guo-Wei Wei}
\email{wei@math.msu.edu}
\affiliation
{Department of Mathematics\\
Michigan State University, MI 48824, USA}
\alsoaffiliation
{Department of Electrical and Computer Engineering\\
Michigan State University, MI 48824, USA}
\alsoaffiliation
{Department of Biochemistry and Molecular Biology\\
Michigan State University, MI 48824, USA}

\title{ The impact of surface area, volume, curvature and Lennard-Jones potential to solvation modeling  }

\begin{document}

%
%
%
%
%

\begin{abstract}
%

This paper explores the impact of  surface area, volume, curvature and Lennard-Jones potential on   solvation free energy predictions. Rigidity surfaces are utilized to generate robust analytical expressions for maximum, minimum, mean and Gaussian curvatures  of solvent-solute interfaces, and define a generalized Poisson-Boltzmann (GPB) equation with a smooth dielectric profile.  Extensive correlation analysis is performed to examine the linear dependence of surface area, surface enclosed volume, maximum curvature, minimum curvature, mean curvature and Gaussian curvature for solvation modeling. It is found that surface area and surfaces enclosed volumes are highly correlated to each others, and poorly correlated to various curvatures for six test sets of molecules. Different curvatures are weakly correlated to each other for six test sets of molecules, but are strongly correlated to each other within each test set of molecules.  Based on correlation analysis, we construct twenty six nontrivial nonpolar solvation models. Our numerical results reveal that the Lennard-Jones (LJ) potential plays a vital role in nonpolar solvation modeling, especially for molecules involving strong van der Waals interactions.  It is found that curvatures are at least as important as surface area or surface enclosed volume in nonpolar solvation modeling.  In conjugation with the GPB model, various curvature based nonpolar solvation models are shown to offer some of the best solvation free energy predictions for a wide range of test sets. For example,  root mean square errors from a  model constituting surface area, volume, mean curvature and LJ potential are less than  0.42 kcal/mol for all test sets.

\end{abstract}
\maketitle



\section{Introduction}
All essential biological processes, such as signaling, transcription, cellular differentiation, etc., take place in an aqueous environment. Therefore, a prerequisite of understanding such biological processes is to study the solvation process, which involves a wide range of solvent-solute interactions, including   hydrogen bonding, ion-dipole, induced dipole, and dipole-dipole,  hydrophobic/hydrophobic, dispersive attractions, or van der Waals forces. The most commonly available experimental measurement of the solvation process is the solvation free energy, i.e., the energy released from the solvation process. As a result, the prediction of solvation free energy has been a main theme of solvation modeling and analysis.  Numerous computational models have been proposed for  solvation free energy prediction, including  molecular mechanics, quantum mechanics, statistical mechanics, integral equation, explicit solvent models, and implicit solvent models \cite{ZhanChen:2010a,ZhanChen:2010b,ZhanChen:2011a}. Each approach has its own advantages, merits and limitations. Among these models, explicit \cite{Ponder:2003a} and quantum methods \cite{Husowitz:2007,Truhlar:2008SolvationP} are ultimately for investigating the solvation of relatively small molecules; however, a great number of degrees of freedom for large systems may lead to unmanageable computational cost. Implicit solvent models, on the contrary, can lower the number of degrees of freedom by approximating the solvent by a continuum representation and describing the solute in atomistic detail \cite{Davis:1990a,Roux:1999,Sharp:1990a}.

In implicit solvent models, the total solvation free energy is divided into nonpolar and polar contributions \cite{Koehl:2006,David:2000}. There is a wide range of implicit solvent models available to describe the polar solvation process; nonetheless, Poisson-Boltzmann (PB) \cite{Baker:2005,Davis:1990a,Fogolari:2002,Sharp:1990a,Zhou:2008b} and generalized Born (GB) models \cite{Bashford:2000,Dominy:1999,Gallicchio:2002,Grant:2007,Onufriev:2002,Tjong:2007b,Tsui:2001} are     commonly used. GB methods are very fast, but are only heuristic models for the polar solvation analysis. PB methods can be derived from fundamental  theories \cite{Beglov:1996,Netz:2000a}; therefore, can offer somewhat of simple   but satisfactorily accurate and robust solvation energy estimations when handling large biomolecules.

To approximate the nonpolar solute-solvent interactions in implicit solvent models, a common way is to assume the nonpolar solvation free energy being correlated with the solvent-accessible surface area (SASA) \cite{Swanson:2004,massova2000combined}, based on the scaled-particle theory (SPT) for nonpolar solutes in aqueous solutions \cite{Stillinger:1973,Pierotti:1976}. However, recent studies indicate that  solvation free energy may depend on both SASA and solvent-accessible volume (SAV), especially in large length scale regimes \cite{Lum:1999,huang2000temperature}. It was pointed out that, unfortunately,  SASA based solvation models do not capture the ubiquitous van der Waals (vdW) interactions near the solvent-solute interface \cite{Gallicchio:2004}. Indeed, the use of SASA, SAV and solvent-solute dispersive interactions to approximate nonpolar energy significantly improves the accuracy of solvation free energy prediction \cite{choudhury2005mechanism,Wagoner:2006,ZhanChen:2012,BaoWang:2015a}. 

One of the most important tasks in handling the implicit solvent models is to define the solute-solvent interface. Many solvation quantities such as surface area, cavitation volume, curvature of the surface  and electrostatic energies significantly depend on the interface definition. The vdW surface, solvent accessible surface \cite{Lee:1971}, and  solvent excluded surface (SES) \cite{Richards:1977} have shown their effectiveness in biomolecular modeling. However, these surface definitions admit geometric singularities \cite{Connolly:1983, Sanner:1996} which result in excessive computational instability and algorithmic effort \cite{Yu:2007,Yu:2007a,Zhou:2006c}. As a result, throughout the past decade, many advanced surface definitions have been developed. One of them is the Gaussian surface description \cite{Grant:1995,MXChen:2011,LiLin:2014}. Another approach is by means of differential geometry. The first curvature induced biomolecular surface was introduced in 2005 using  geometric partial differential equations (PDEs)    \cite{Wei:2005}. The first variational molecular surface based on minimal surface theory was proposed in 2006 \cite{Bates:2006,Bates:2008}.   These surface definitions lead to curvature controlled smooth solvent-solute interfaces that enable one to generate a smooth dielectric profile over  solvent and solute domains. This development leads to differential geometry based solvation models \cite{ZhanChen:2010a,ZhanChen:2010b} and multiscale models \cite{Wei:2009,Wei:2012,Wei:2013}. These models have been confirmed to deliver excellent solvation free energy predictions \cite{ZhanChen:2012,BaoWang:2015a}.
Recently, a family of rigidity surfaces has been proposed in the flexibility-rigidity index (FRI) method, which significantly outperforms the Gaussian network model (GNM) and anisotropic network model  (ANM) in protein B-factor prediction \cite{KLXia:2013d,Opron:2014,Opron:2015a, KLXia:2015f}. Flexibility is an intrinsic property of proteins and is known to be important for protein drug binding \cite{Alvarez-Garcia:2014}, allosteric signaling \cite{ZBu:2011} and self-assembly \cite{Marsh:2014}. It must play an important role in the solvation process because of entropy effects. Therefore, FRI based rigidity surfaces, which can be regarded as generalizations of   classic Gaussian surfaces \cite{Grant:1995,MXChen:2011,LiLin:2014}, may have an advantage in solvation analysis as well.    

 In molecular biophysics,  curvature measures the variability or non-flatness of a biomolecular surface and is believed to play an important role in many biological processes, such as  membrane curvature sensing, and protein-membrane and protein DNA interactions. These interactions may be described by the Canham-Helfrich curvature energy functional \cite{Helfrich:1973}. Due to its potential contribution to the cavitation cost, curvature of the solute-solvent surface  is believed to affect the solvation free energy \cite{Dzubiella:2006}. By using SPT, the surface  tension is assumed to have  a Gaussian curvature dependence \cite{Dzubiella:2006}. The curvature in such cases is locally estimated and is a function of the solvent radius. Nevertheless, the quantitative contribution of various curvatures to solvation free energy prediction has not been investigated.   

The objective of the present work is to explore the impact of surface area, volume, curvature, and Lennard-Jones potential on the solvation free energy prediction. We are particularly interested in the role of Hadwiger integrals, namely area, volume, Gaussian curvature and mean curvature, to  the molecular solvation analysis.  Therefore, we consider Gaussian curvature and  mean curvature, as well as minimum and maximum curvatures in the present work. For the sake of accurate and analytical curvature estimation,  we employ rigidity surfaces that not admit geometric singularities. Unlike the geometric flow surface  in our previous work \cite{ZhanChen:2010a,BaoWang:2015a}, the construction of rigidity surfaces does not require a surface evolution; accordingly, does not need parameter constraints to stabilize the optimization process. In the current models, instead of local curvature considered in other work \cite{sharp1991extracting,jackson1994application,Dzubiella:2006},  total curvatures that are the summations of absolute local curvatures  are employed to measure the total variability of solvent-solute interfaces. We show that curvature based nonpolar solvation models offer some of the best solvation predictions for a large amount of molecules.   

The rest of this paper is organized as follows. Section \ref{sec.theory_models} presents the theory and formulation of new solvation models. We first briefly introduce the rigidity surface for the surface definition. A generalized PB equation using a smooth dielectric function is formulated.  We provide an advanced algorithm for the evaluation of surface area and surface enclosed volume. Analytical presentation for calculating various curvatures, namely Gaussian curvature, mean curvature, minimum and maximum principal curvatures are presented. Finally, we  introduces a parameter learning algorithm to solvation energy prediction. Section \ref{sec.results} is devoted to numerical studies. First, we discuss the dataset used in this work. Over a hundred molecules of both polar and nonpolar types are employed in our numerical tests. We then discuss the models and their  abbreviations to be used in this study. The numerical setups for  nonpolar and polar solvation free energy calculations are described in detail. We  explore the correlations between area, volume, and different types of curvatures. Based on the  root mean square error (RMSE) computed between experimental and predicted results, we reveal the impact of each interested nonpolar quantities on solvation free energy prediction. The final part of Section \ref{sec.results} is devoted to the investigation of the most accurate and reliable solvation model. This paper ends with a conclusion.

\section{ Models and algorithms}\label{sec.theory_models}

\subsection{Solvation models}\label{sec.solvation_model}

The solvation free energy, $\Delta G$,  is calculated as a sum of polar, $\Delta G^\text{p}$, and nonpolar, $G^\text{np}$, components
\begin{align}\label{solvation_eng}
\Delta G = \Delta G^\text{p} + G^\text{np}.
\end{align}
Here,  $\Delta G^\text{p}$ is modeled by the Poisson-Boltzmann theory.  For the nonpolar contribution, we consider the following nonpolar solvation free functional
\begin{align} 
\Delta G^\text{np}=\gamma A + pV + \sum_j\lambda_j C_j +\rho_0 \int_{\Omega_s} U^{\text{vdW}}\mathrm{d}\mathbf{r},
\label{gnp}
\end{align}
where $A$ and $V$ are, respectively, the surface area and surface enclosed volume of  the solute molecule of interest. Additionally, $\gamma$ is the surface tension and $p$ is  the hydrodynamic pressure difference. We denote  $C_j$ and $\lambda_j$ respectively   curvatures and associated bending coefficients of the molecular surface. Thus, the index $j$ runs from  maximum curvature, minimum curvature,      mean curvature to  Gaussian curvature. Here $\rho_0$ is the solvent bulk density, and $U^\text{vdW}$ is the  van der Waals (vdW) interaction approximated by the Lennard-Jones potential. The final integral is  computed solely over solvent domain  $\Omega_s$. One can turn off certain terms in Eq. (\ref{gnp}) to arrive at simplified models.

\subsection{Rigidity surface}\label{sec.rig_surface}

	Flexibility-rigidity index (FRI)  has been shown to significantly outperform other methods, such the Gaussian network model (GNM) and anisotropic network model  (ANM), in protein flexibility analysis or B-factor prediction  over hundreds of molecules \cite{KLXia:2013d,Opron:2014,Opron:2015a, KLXia:2015f}. 
	Given a  molecule with $N$ atoms, we denote $\mathbf{r}_j$ the position of $j$th atom, $\|\mathbf{r} - \mathbf{r}_j \|$ the Euclidean distance between a point $\mathbf{r}$ and atom $\mathbf{r}_j$. In our FRI method,  commonly used correlation kernels or statistical density estimators  \cite{KLXia:2013d,Opron:2014,GWei:2000} include generalized exponential functions
	\begin{align} \label{exponent}
	\mathbf{\Phi}\left(\|\mathbf{r}-\mathbf{r}_j\|;\eta_j\right) = e^{{-\left(\|\mathbf{r}-\mathbf{r}_j\|/\eta_j\right)}^{\kappa}}, \quad \kappa > 0,
	\end{align}
	and generalized Lorentz functions
	\begin{align}\label{lorentz}
	\mathbf{\Phi}\left(\|\mathbf{r}-\mathbf{r}_j\|;\eta_j\right)=\frac{1}{1 +\left(\frac{\|\mathbf{r}-\mathbf{r}_j\|}{\eta_j}\right)^{\nu}},\quad \nu > 0,
	\end{align}
	where $\eta_j$ is a scale parameter. 
	An atomic rigidity function $\mu(\mathbf{r})$ for  an arbitrary point $\mathbf{r}$ on the computational domain can be defined as 
	\begin{align}
	\mu(\mathbf{r})=\sum_{j=1}^{N} w_j (\mathbf{r})\mathbf{\Phi}\left(\|\mathbf{r}-\mathbf{r}_j\|;\eta_j\right),\label{total_density_1}
	\end{align}
	where $ w_j (\mathbf{r})$ is a  weight function. The atomic rigidity function $\mu(\mathbf{r})$ measures the atomic density at position $\mathbf{r}$. This intepretation can be easily verified since if we choose  $ w_j (\mathbf{r})$ such that 
	$$
	\int	\mu(\mathbf{r})d{\bf r}=1.
	$$
	Then the	atomic rigidity function $\mu(\mathbf{r})$ becomes a  probability density distribution such that $\mu(\mathbf{r})d{\bf r}$ is the probability of finding all the $N$ atoms in an infinitesimal volume element $d{\bf r}$ at a given point ${\bf r}\in \mathbb{R}^3$.  For $\mathbf{\Phi}\left(\|\mathbf{r}-\mathbf{r}_j\|;\eta_j\right) = e^{{-\left(\|\mathbf{r}-\mathbf{r}_j\|/\eta_j\right)}^2}$, one can analytically choose $w_j (\mathbf{r})= \frac{1}{N} \left(\frac{1}{\pi \eta_j^2} \right)^{\frac{3}{2}}$ to normalize     	atomic rigidity function  $\mu(\mathbf{r})$.

	For simplicity, in this work we just employ the Gaussian kernel, i.e., generalized exponential kernel with $\kappa=2$, $\eta_j=r^{\text{vdW}}_j$ (i.e., the vdW radius of atom $j$), and $w_j=1$ for all $j=1,2,\cdots,N$. Other FRI kernels are found to deliver very similar results. Our rigidity surfaces can be regarded as a generalization of Gaussian surfaces  \cite{Zap,Grant:2007}.
	
\subsection{Smooth rigidity function-based dielectric function}\label{sec.smooth_die}

We denote $\Omega$ the total domain, and $\Omega$ is divided into two regions, i.e., aqueous solvent domain $\Omega_s$ and  solute molecular domain $\Omega_m$.
Our ultimate goal is to construct a smooth dielectric function in a similar way to that of differential geometry based solvation models as follows \cite{Wei:2009,ZhanChen:2010a,ZhanChen:2010b} 
\begin{align}\label{dielectric_function_1}
\epsilon(\mu)=(1-\mu)\epsilon_s + \mu \epsilon_m,
\end{align}
where $\epsilon_s$ and $\epsilon_m$ are the dielectric constants of the solvent and solute, respectively.  However the total atomic density described in \eqref{total_density_1} exceeds 1 in many cases. As a result, we normalize the atomic rigidity function as
\begin{align}
\bar{\mu}(\mathbf{r})=\frac{1}{\max\limits_{\mathbf{r}\in\Omega} \mu(\mathbf{r})}\mu(\mathbf{r}).
\end{align}
Nonetheless, the dielectric function \eqref{dielectric_function_1} is still not applicable since the characteristic function $1-\bar{\mu}$  may not  capture the commonly defined solvent domain. This is due to the fact that the value of $\bar{\mu}(\mathbf{r})$ could be less than 1 inside the biomolecule. As a result, we define the molecular domain as $\{\mathbf{r}\in\Omega | \mu(\mathbf{r}) \geq \beta\}$, where $\beta$ is a cut-off value defined in the protocol to attain the best fitting against other PB solvers, such as MIBPB \cite{DuanChen:2011a}. By doing so, the dielectric function \eqref{dielectric_function_1} will be modified as the following
\begin{align}
\epsilon(\bar{\mu}(\mathbf{r}))=\left\{
\begin{array}{ll}
\epsilon_m, & \text{if $\bar{\mu}(\mathbf{r})\geq\beta$,} \\
\displaystyle\left(1-\frac{\bar{\mu}}{\beta}\right)\epsilon_s + \frac{\bar{\mu}}{\beta}\epsilon_m, &  \text{if $\bar{\mu}(\mathbf{r})<\beta$.}
\end{array}
\label{dielectric_funtion_2}
\right.
\end{align}

\subsection{Generalized Poisson-Boltzmann (GPB) equation}\label{sec.gpb}

With smooth dielectric profile being defined in \eqref{dielectric_funtion_2}, we arrive at the GPB equation in an ion-free     solvent
\begin{align}\label{gpb}
-\nabla\cdot\left(\epsilon(\bar{\mu})\nabla\phi({\bf r})\right)= \bar{\mu}\rho_m({\bf r}),
\end{align} 
where $\phi$ is the electrostatic potential, $\rho_m({\bf r})=\sum_i^{N_m}Q_i\delta({\bf r-r}_i)$ represents the fixed charge density of the solute. Here $Q(\mathbf{r}_i)$ is the partial charge at $\mathbf{r_i}$ in the solute molecule, and $N_m$ is the total number of partial charges. 

Let $\Omega$ be the computational domain of the GPB equation. Without considering the salt molecule in the solvent, we employ the Dirichlet boundary condition via a Debye-H\"{u}ckel expression for the GPB equation
\begin{align}
\phi(\mathbf{r})=\sum_{i=1}^{N_m}\frac{Q_i}{\epsilon_s\|\mathbf{r}-\mathbf{r}_i\|} ,\quad \forall\mathbf{r}\in\partial\Omega.
\end{align} 

The electrostatic solvation free energy, $\Delta G^{\text{p}}$, is calculated by
\begin{align}
\Delta G^\text{p} = \frac{1}{2}\sum_{i=1}^{N_m}Q(\mathbf{r}_i)\left(\phi(\mathbf{r}_i)-\phi_0(\mathbf{r}_i)\right),
\end{align}
where $\phi$ and $\phi_0$ are, respectively, the electrostatic potential in the presence of the solvent and vacuum. In other words, $\phi$ is a solution of the GPB equation \eqref{gpb}, and homogeneous solution $\phi_0$ of the GPB equation is obtained by setting dielectric function $\epsilon(\bar{\mu})=\epsilon_m$ in the whole computational domain $\Omega$. 

	\subsection{Surface area and surface-enclosed volume}\label{sec.area_volume}
	
	The surface integral for a density function $f$ over $\Gamma$ in the domain $\Omega$ with a uniform mesh can be evaluated by
	\cite{Geng:2011,QZheng:2012,WFTian:2014}
	\begin{align}
	\int_{\Gamma} f(x,y,z)dS \approx \sum_{(i,j,k)\in I}\left(f(x_0,y_j,z_k)\frac{|n_x|}{h} + f(x_i,y_0,z_k)\frac{|n_y|}{h} + f(x_i,y_j,z_0)\frac{|n_z|}{h}\right)h^3,\label{area}
	\end{align}
	where $(x_0,y_j,z_k)$ is the intersecting point between the interface $\Gamma$ and the $x$ mesh line going through $(i,j,k)$, and $n_x$ is the $x$ component of the unit normal vector at $(x_0,y_j,z_k)$. Similar definitions are used for the $y$ and $z$ directions. We only carry out the calculation \eqref{area} in a small set of irregular grid points, denoted as $I$. Here, the irregular grid points are defined to be the points associated with neighbor point(s) from the other side of the interface $\Gamma$ in the second order finite difference scheme \cite{Yu:2007}. In this case, $I$ will contain the irregular points near  interface $\Gamma$. Finally, $h$ is the uniform grid spacing. The volume integral can be simply approximated by
	\begin{align}
	\int_{\Omega_m} f d\mathbf{r} \approx \sum_{(i,j,k)\in J}f(x_i,y_j,z_k)h^3,\label{volume}
	\end{align}
	where $\Omega_m$ is the domain enclosed by $\Gamma$, and $J$ is the set of all grid points inside $\Omega_m$. By considering the density function $f=1$, Eqs. \eqref{area} and \eqref{volume} can be respectively used for the surface area and volume calculations.
	
	\subsection{Curvature calculation}\label{sec.curve}
	
	The evaluation of the curvatures for isosurface embedded volumetric data, $S(x,y,z)$, has been reported in the literature \cite{Bates:2008,Soldea:2006,KLXia:2014a}. In general, there are two approaches for the curvature evaluation. The first method is to invoke   the first and second fundamental forms in differential geometry, the another one is to make use of the Hessian matrix method \cite{Kindlmann:2003}. Since both of these algorithms yield the same results as shown in our earlier work  \cite{KLXia:2014a}, only the first approach is employed in the present work. To this end, we immediately provide the formulation for Gaussian curvature ($K$) and mean curvature ($H$) by means  of the  first and second fundamental forms \cite{Soldea:2006,KLXia:2014a}
	\begin{align}
	K=&\frac{2S_x S_y S_{xz}S_{yz} + 2S_xS_zS_{xy}S_{yz}+2S_yS_zS_{xy}S_{xz}}{g^2}\nonumber\\
	&-\frac{2 S_x S_z S_{xz} S_{yy} + 2 S_y S_z S_{xx} S_{yz} + 2 S_x S_y S_{xy} S_{zz}}{g^2}\nonumber\\
	&+\frac{S_z^2 S_{xx}  S_{yy} + S_x^2 S_{yy} S_{zz} + S_y^2 S_{xx} S_{zz}}{g^2}\nonumber\\
	&-\frac{S_x^2 S_{yz}^2 + S_y^2 S_{xz}^2 + S_z^2 S_{xy}^2}{g^2},\label{gaussian_curv}
	\end{align}
	and
	\begin{align}
	H=\frac{2 S_x S_y S_{xy} + 2 S_x S_z S_{xz} + 2 S_y S_z S_{yz} - (S_y^2 + S_z^2)S_{xx} - (S_x^2 + S_z^2)S_{yy} - (S_x^2+S_y^2)S_{zz}}{2g^{\frac{3}{2}}},
	\label{mean_curv}
	\end{align}
	where $g=S_x^2 + S_y^2 + S_z^2$.
	With determined Gaussian and mean curvatures, the minimum, $\kappa_1$, and maximum, $\kappa_2$, can be evaluated by
	\begin{align}
	\kappa_1=\min\{H-\sqrt{H^2-K},H+\sqrt{H^2-K}\},\quad \kappa_2=\max\{H-\sqrt{H^2-K},H+\sqrt{H^2-K}\}.\label{minmax_curv}
	\end{align}
	We    apply the formulations \eqref{gaussian_curv}, \eqref{mean_curv} and \eqref{minmax_curv} for curvature calculations of rigidity surfaces. Again, we only consider generalized exponential kernel with $\kappa=2$ and $w_j=1$ for all $j=1,2,\cdot,N$ in this paper. As a result, the atomic rigidity function $\mu(\mathbf{r})$, defined in \eqref{exponent} and \eqref{total_density_1},  become
	\begin{align}\label{gaussian_surface}
	\mu(\mathbf{r}) = \sum_{j=1}^{N}e^{-\left(\frac{\| r-r_j\|}{\eta_j}\right)^2} =\sum_{j=1}^{N}e^{-\frac{(x-x_j)^2+(y-y_j)^2+(z-z_j)^2}{\eta_j^2}}.
	\end{align}
	
	Note that derivatives of $\mu$ can be analytically attained. Therefore,  	by replacing $S$ with $\mu$ in various curvature formulas, we  obtain  analytical expressions for different curvatures of FRI based rigidity surfaces. As a result, the calculation of various curvatures is very simple and robust for rigidity surfaces.

	\subsection{Optimization algorithm}\label{sec.opt_alg}
	
	In this section, we  present an algorithm, inspired by the algorithm 2 in our earlier work \cite{BaoWang:2015a}, to optimize the parameters appearing in the nonpolar component. In this work, we utilize the 12-6 Lennard-Jones potential to model the van der Waals interaction $U_i^{\text{vdW}}$ regarding an atom of type $i$
	\begin{align}
	U_i^{\text{vdW}}(\mathbf{r})=\epsilon_i \left[\left(\frac{\sigma_i+\sigma_s}{\|\mathbf{r}-\mathbf{r}_i\|}\right)^{12} - 2 \left(\frac{\sigma_i+\sigma_s}{\|\mathbf{r}-\mathbf{r}_i\|}\right)^6\right],
	\end{align}
	where $\epsilon_i$ is the well-depth parameter, $\sigma_i$ and $\sigma_s$ are, respectively, the radii of the atom of type $i$ and solvent. Here $\mathbf{r}$ is the location of an arbitrary point in the solvent domain, and $\mathbf{r}_i$ is the location of the atom of type $i$. Since the integral of the Lennard-Jones potential term involves in the solvent bulk density $\rho_0$, the fitting parameter for the van der Waals interaction of the atom of type $i$ will be $\tilde{\epsilon}_i  \doteq \rho_0\epsilon_i$.
	Assume that we have a training group containing $n$ molecules, the process of calculating solvation free energy 
	will give us the following quantities for the $j$th $(j=1,2,\cdots,n)$ molecule
	\begin{align}
	\left\{\Delta G^{\text{p}}_j,A_j,V_j,C_{1j},C_{2j},C_{3j},C_{4j},\left(\sum_{i=1}^{N_m}\delta^1_i\int_{\Omega_s}U_1^{\text{vdW}}(\mathbf{r})\mathrm{d}\mathbf{r}\right)_j,\cdots,
	\left(\sum_{i=1}^{N_m}\delta^{N_t}_i\int_{\Omega_s}U_{N_t}^{\text{vdW}}(\mathbf{r})\mathrm{d}\mathbf{r}\right)_j
	\right\},
	\end{align}
	where $N_m$ and $N_t$ are the number of atoms and the number of atom types in each individual molecule, respectively and $C_{ij}$ denotes the $i$th curvature for the $j$th molecule. Here $\delta^k_i$ is defined as follows
	\begin{align}
	\delta^k_i=\left\{\begin{array}{ll}
	1,& \text{if atom $i$ belongs to type $k$,}\\
	0,& \text{otherwise,}
	\end{array}
	\right.
	\end{align}
	where $k=1,2,\cdots,N_t$ and $i=1,2,\cdots,N_m$. We denote the parameter set for the current training group as $\mathbf{P}=\left\{\gamma,p,\lambda_1,\cdots,\lambda_4,\tilde{\epsilon}_1,\tilde{\epsilon}_2,\cdots,\tilde{\epsilon}_{N_t}\right\}$. The solvation free energy for molecule $j$ will be then predicted by
	\begin{align}
	\Delta G_j =&\Delta G^{\text{p}}_j+\gamma A_j + p V_j + \sum_i\lambda_i C_{ij} + \tilde{\epsilon}_1 \left(\sum_{i=1}^{N_m}\sigma^1_i\int_{\Omega_s}U_1^{\text{vdW}}(\mathbf{r})\mathrm{d}\mathbf{r}\right)_j \nonumber\\ 
	&+ \cdots +
	\tilde{\epsilon}_{N_t}\left(\sum_{i=1}^{N_m}\sigma^{N_t}_i\int_{\Omega_s}U_{N_t}^{\text{vdW}}(\mathbf{r})\mathrm{d}\mathbf{r}\right)_j.
	\label{solvation_param}
	\end{align}
	It is noted that the fitting parameter of corresponding vanishing term will set to $0$ in the solvation free energy calculation \eqref{solvation_param}. We denote a vector of predicted solvation energies for the given molecular group as $\Delta\mathbf{G}(\mathbf{P})=(\Delta G_1,\Delta G_2,\cdots,\Delta G_n)$ which depends on the parameter set $\mathbf{P}$. In addition, we denote a vector of the corresponding experimental solvation free energy as $\Delta\mathbf{G}^{\text{Exp}}=(\Delta G^{\text{Exp}}_1,\Delta G^{\text{Exp}}_2,\cdots,\Delta G^{\text{Exp}}_n)$. We then optimize the parameter set $\mathbf{P}$ by solving the following minimization problem
	\begin{align}
	\min_{\mathbf{P}}\left(\|\Delta \mathbf{G}(\mathbf{P}) - \Delta\mathbf{G}^{\text{Exp}}\|_2\right),\label{min_problem}
	\end{align}
	where $\|*\|_2$ denotes the $L_2$ norm of the quantity $*$. Optimization problem \eqref{min_problem} is a standard one which can be solved by many available tools. In this work, we employ CVX software \cite{cvx}  to deal with it.
	
	Unlike our previous work \cite{BaoWang:2015a}, we only need to generate the fixed molecular surface and solve the GPB equation \eqref{gpb} one time. We will then utilize the optimization process \eqref{min_problem} with obtained quantities to achieve the optimized parameter set $\mathbf{P}$.

\section{Results and discussions} \label{sec.results}

\subsection{Data sets}\label{sec:DataSet} 

To study the impact of area, volume, curvature and Lennard-Jones potential on the solvation free energy prediction, we employ a large number of solute molecules with accurate experimental solvation values. These molecules  are of both  polar and nonpolar types and are divided into six groups: the SAMPL0 test set \cite{Nicholls:2008solvation} with 17 molecules, alkane set with 35 molecules, alkene set with 19 molecules, ether set with 15 molecules, alcohol set with 23 molecules, and phenol set with 18 molecules sets \cite{Mobley:2014}. The charges of the SAMPL0 set are taken from the OpenEye-AM1-BCC v1 parameters \cite{Jakalian:2000}, while their atomic coordinates and radii are based on the ZAP-9 parametrization \cite{Nicholls:2008solvation}. The structural conformations for the other groups are adopted from FreeSolv\cite{Mobley:2014} with their parameter and coordinate information being downloaded from Mobley's homepage \url{http://mobleylab.org/resources.html}.

\subsection{Model abbreviation}\label{sec:classification} 
\begin{table}[!ht]
	\centering
	\caption{Model terminologies}
	\label{tab.symbol}
	\begin{tabular}{cl}
		Symbols & Meaning\\
		\hline
		$\mathbf{A}$	    & 	$G^\text{np}$ contains a area term \\
		$\mathbf{V}$	   	&   $G^\text{np}$ contains a volume term \\
		$\mathbf{L}$		  & 	$G^\text{np}$ contains a Lennard-Jones potential term\\
		$\mathbf{k_1}$		&	  $G^\text{np}$ contains a minimum curvature term\\
		$\mathbf{k_2}$		&	  $G^\text{np}$ contains a maximum curvature term\\
		$\mathbf{H}$		  &	  $G^\text{np}$ contains a mean curvature term\\
		$\mathbf{K}$		  &	  $G^\text{np}$ contains a Gaussian curvature term\\
		\hline
	\end{tabular}
\end{table}

It is noted that if we only consider area, volume and van der Waals interaction in  nonpolar component computations, we would arrive at the formulation already discussed in the literature \cite{Wagoner:2006,ZhanChen:2010a}. However, the nonpolar component in this work includes additional  curvature terms. To investigate the impact of  area, volume, Lennard-Jones potential and curvature on the   solvation free energy prediction, we benchmark different models consisting of various terms in nonpolar free energy functionals. To this end, we use the symbols listed in Table \ref{tab.symbol} to label a model if it includes the corresponding terms in the nonpolar solvation free functional. For example, model  $\mathbf{A}$ only considers the surface area term, whereas model $\mathbf{AVL}$   incorporates area ({\bf A}), volume ({\bf V}) and Lennard-Jones potential ({\bf L}) terms in nonpolar energy calculations.

\subsection{Polar and nonpolar calculations}

In this work, we employ   rigidity surface \cite{KLXia:2013d,Opron:2014}, discussed in Section \ref{sec.rig_surface}, as the   surface representation of a solvent-solute interface. For simplicity, we implement the Gaussian kernel for all tests, while other FRI kernels deliver similar results. 

\paragraph{Polar part} 
By following the paradigm for constructing a smooth dielectric function in differential geometry based solvation models \cite{Wei:2009,ZhanChen:2010a}, we propose a smooth rigidity-based dielectric function as in Eq. \eqref{dielectric_funtion_2}.
The generalized Poisson-Boltzmann (GPB) equation described in Eq. \eqref{gpb}
is used. For the current framework, we consider the solvent environment without salt and there is only one solvent component, water. The polar solvation energy is then calculated as the  difference of the GPB energies in water and in a vacuum, and the detail of this representation is offered in Section \ref{sec.gpb}. Similar results are obtained if we create a sharp interface and then employ a standard PB solver to compute the polar solvation energy.

In all calculations, the rigidity surface is constructed based on the cut-off value being $\beta =0.09$, and the dielectric constants for solute  and solvent regions are set to 1 and 80, respectively. In addition, the grid spacing is set to $0.2$ \AA. The computational domain is the bounding box of the molecular surface with an extra buffer length of $3$ \AA. The changes in RMS errors are less than 0.02 kcal/mol when the buffer length is extended to 6 \AA.
Since the dielectric profile in the GPB equation is smooth throughout the computational domain, one can easily make use of the standard second order finite difference scheme to numerically solve the GPB equation. Then, a standard Krylov subspace method based solver \cite{ZhanChen:2010a,ZhanChen:2010b} is employed to handle the resulting algebraic equation system.

\paragraph{Nonpolar part}

To estimate the surface area and surface enclosed volume for a rigidity surface, we utilize a stand-alone algorithm based on the marching cubes method, and the detail of this procedure is referred to Section \ref{sec.area_volume}. Thanks to the use of the rigidity surface, the curvature of a solvent-solute interface can be analytically determined instead of using numerical approximations as in our earlier   differential geometry model  \cite{KLXia:2014a}. To prevent the   curvature from canceling each other   at different grid points, we construct   total  curvatures defined as
\begin{align}
C_j=\sum_{\mathbf{r}_i\in I} |c_j(\mathbf{r}_i)|h^2,
\end{align}
where $\mathbf{r}_i$ is the position of the $i$th  grid point, $I$ is a set of irregular grid points in the region of the solvent-solute boundary \cite{Yu:2007,Yu:2007a,Zhou:2006c} and $h$ is the mesh size of the uniform computational domain. Here $c_j(\mathbf{r}_i)$ is the $j$th type of curvature at position $\mathbf{r}_i$, and  index $j$ runs through  minimum, maximum, mean and Gaussian curvatures. Since the full standard 12-6 Lennard-Jones potential improves accuracy of the solvation free energy prediction \cite{ZhanChen:2011a,BaoWang:2015a}, it is utilized to model the    vdW  interaction $U^{\text{vdW}}$ in the current work.

	\begin{figure}[!tb]
		\begin{center}
			\includegraphics[width=0.70\columnwidth]{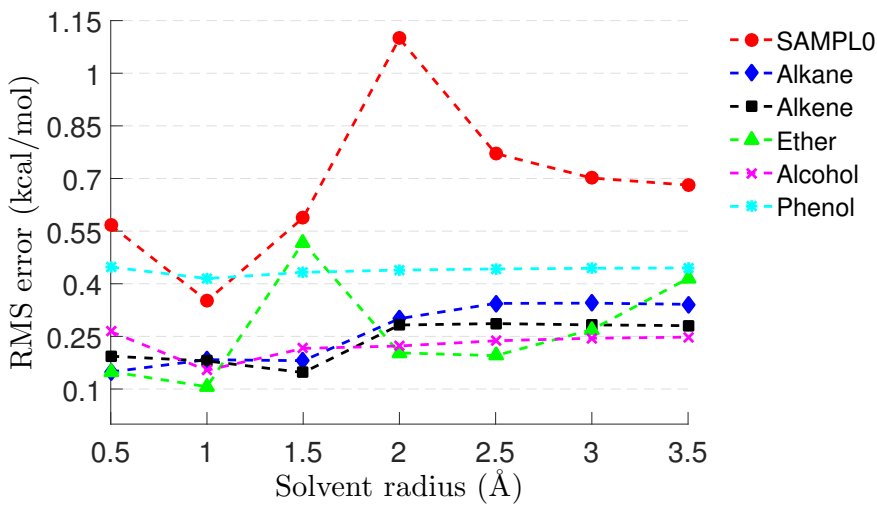}
			\caption{The relations between the solvent radii and the RMS errors for model $\mathbf{AVHL}$. Red circle: SAMPL0   set; blue diamond: alkane set; black square: alkene set; green triangle: ether set ; pink cross: alcohol set; cyan asterisk: phenol set.  }
			\label{fig.optimal_rad}
		\end{center}
	\end{figure}

Similar to our previous work \cite{BaoWang:2015a}, an optimization process as discussed in Section \ref{sec.opt_alg} is applied
 to determine the optimal parameters for the nonpolar free energy calculations.  Unfortunately, the involvement of the solvent radius in the Lennard-Jones potential term features a high nonlinearity. Consequently, it cannot be incorporated into the parameter optimization. Instead, we resort to a brute force approach to determine the most favorable solvent radius for six molecular sets including SAMPL0, alkane, alkene, ether, alcohol, and phenol groups. The value of $\sigma_s$ that mostly produces the smallest RMS error between predicted and experimental solvation free energies will be employed in all numerical calculations. 	By considering model $\mathbf{AVHL}$, we depict the relations between RMS errors and the solvent radii varying from $0.5$ \AA $ $ to $3.5$ \AA $ $ with the increment of $0.5$ \AA $ $ in Fig. \ref{fig.optimal_rad}. This figure reveals that the use of $\sigma_s=1$ \AA $ $  will give us the smallest RMS errors in all test sets except alkane and alkene sets. Therefore, we utilize solvent radius $1$ \AA $ $ for the current work.

\subsection{Correlations between area, volume and curvatures} \label{sec.correl}

Understanding the correlation or   non-correlation between different modeling components is important for analyzing solvation models. A strong correlation between any pair of components indicates their strong linear dependence and redundancy in optimization based solvation modeling. While a weak correlation implies their  complementary roles in an optimization based  solvation modeling.  

\begin{figure}[!ht]
	\begin{center}
		\includegraphics[width=0.50\columnwidth]{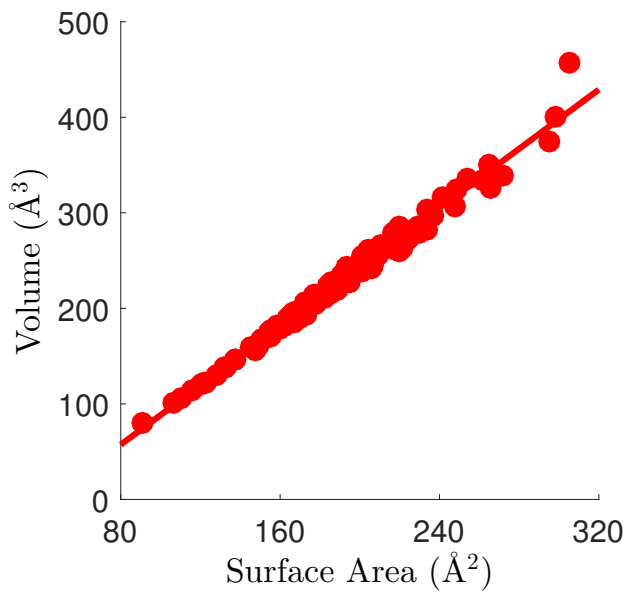}\quad
		\caption{Area versus volume over 127 molecules in all six groups. $R^2=0.99$, and fitting line: $y=1.55x-66.51$.}
		\label{fig.correl_area_volume}
	\end{center}
\end{figure} 

\begin{figure}[!ht]
	\begin{center}
		\includegraphics[width=0.45\columnwidth]{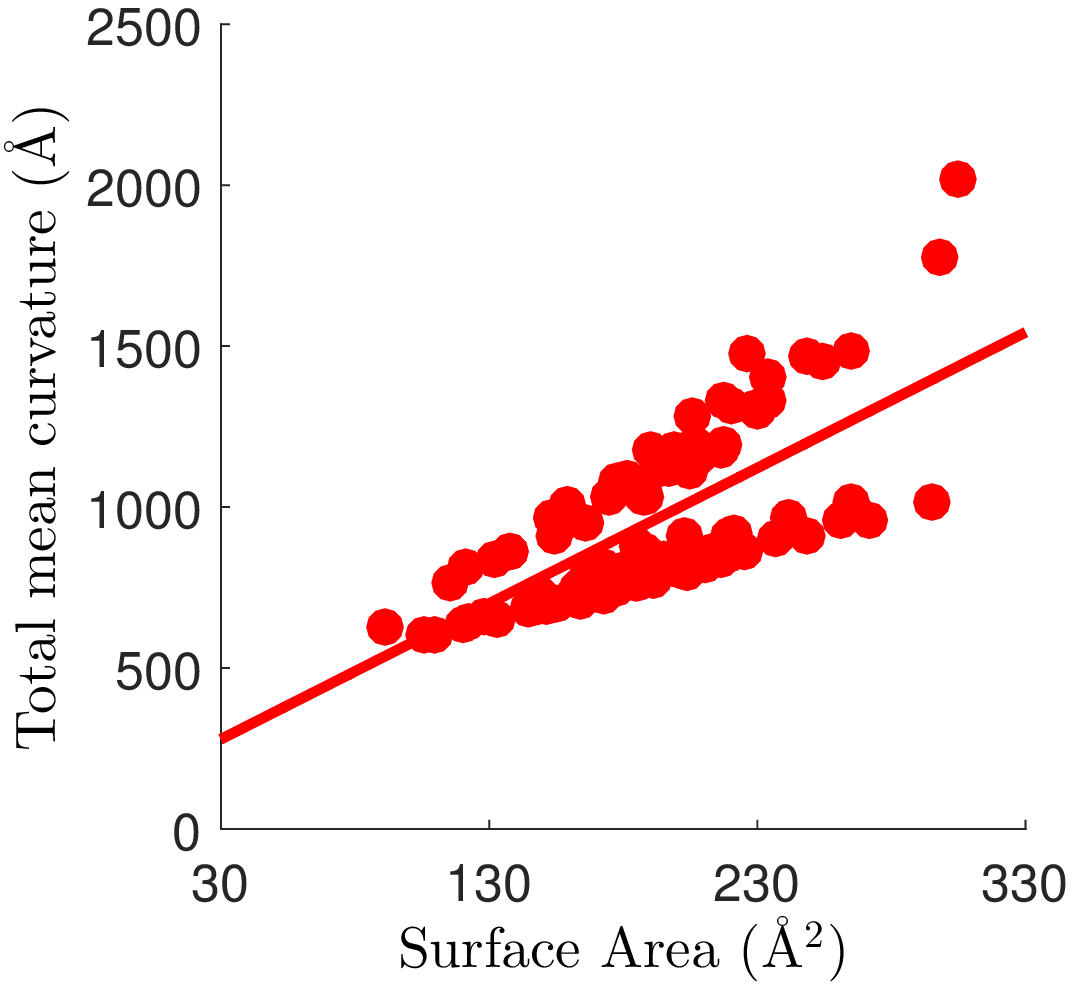}\quad
		\includegraphics[width=0.45\columnwidth]{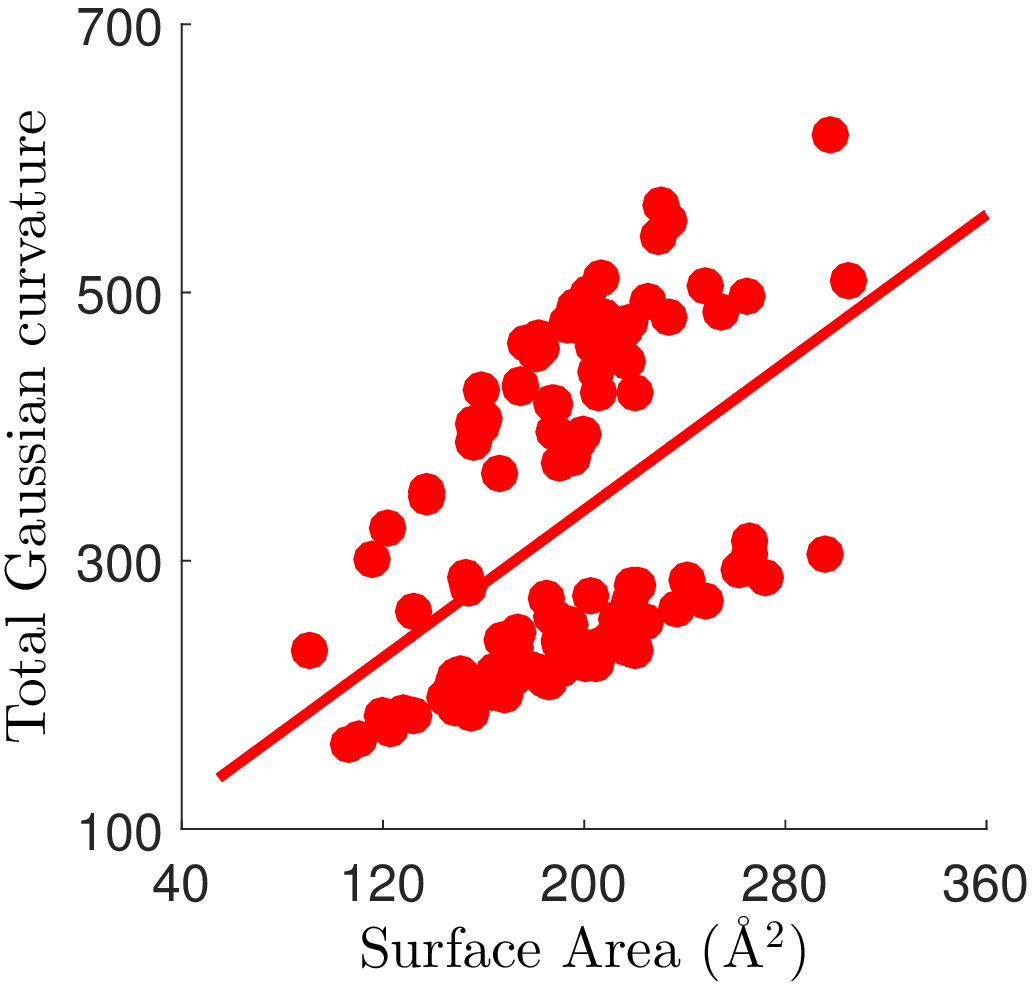}\\
		\vspace*{0.5cm}
		\includegraphics[width=0.45\columnwidth]{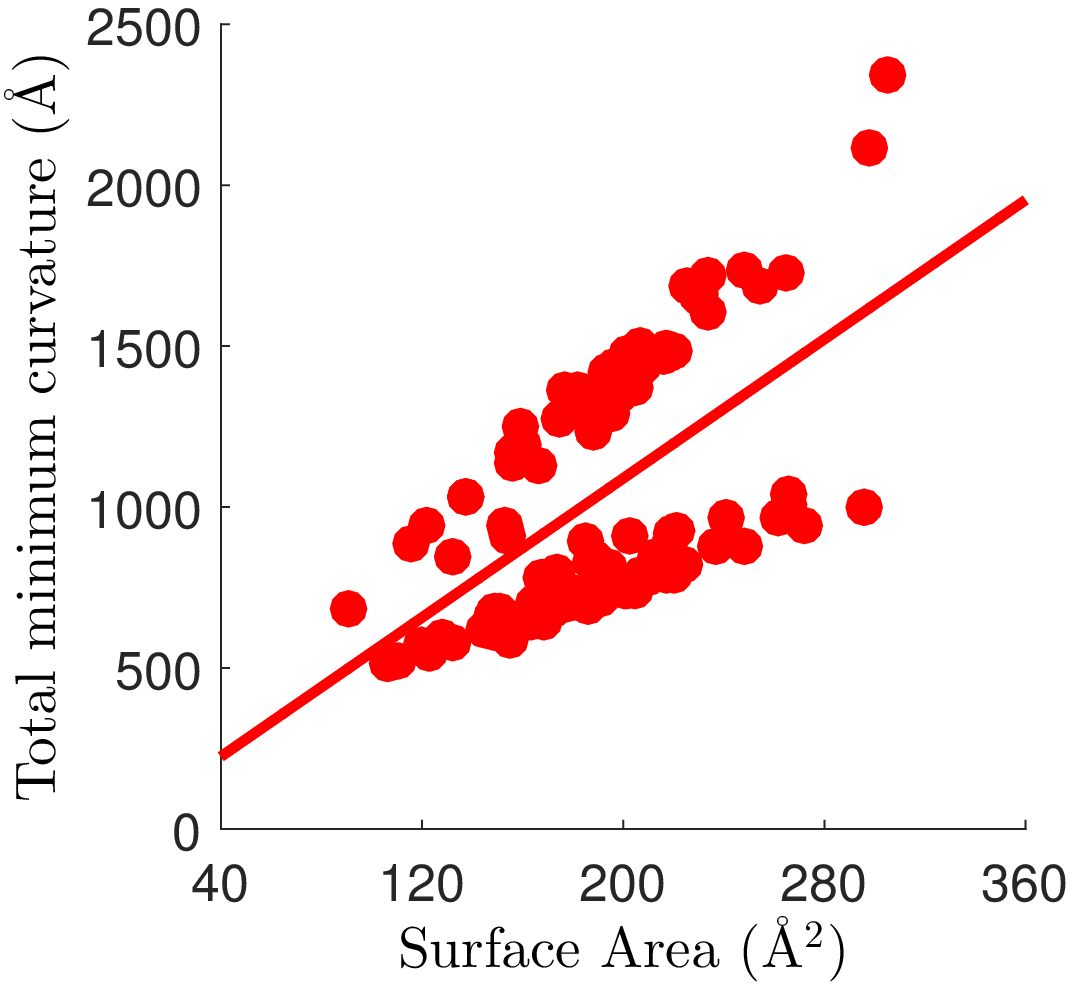}\quad
		\includegraphics[width=0.45\columnwidth]{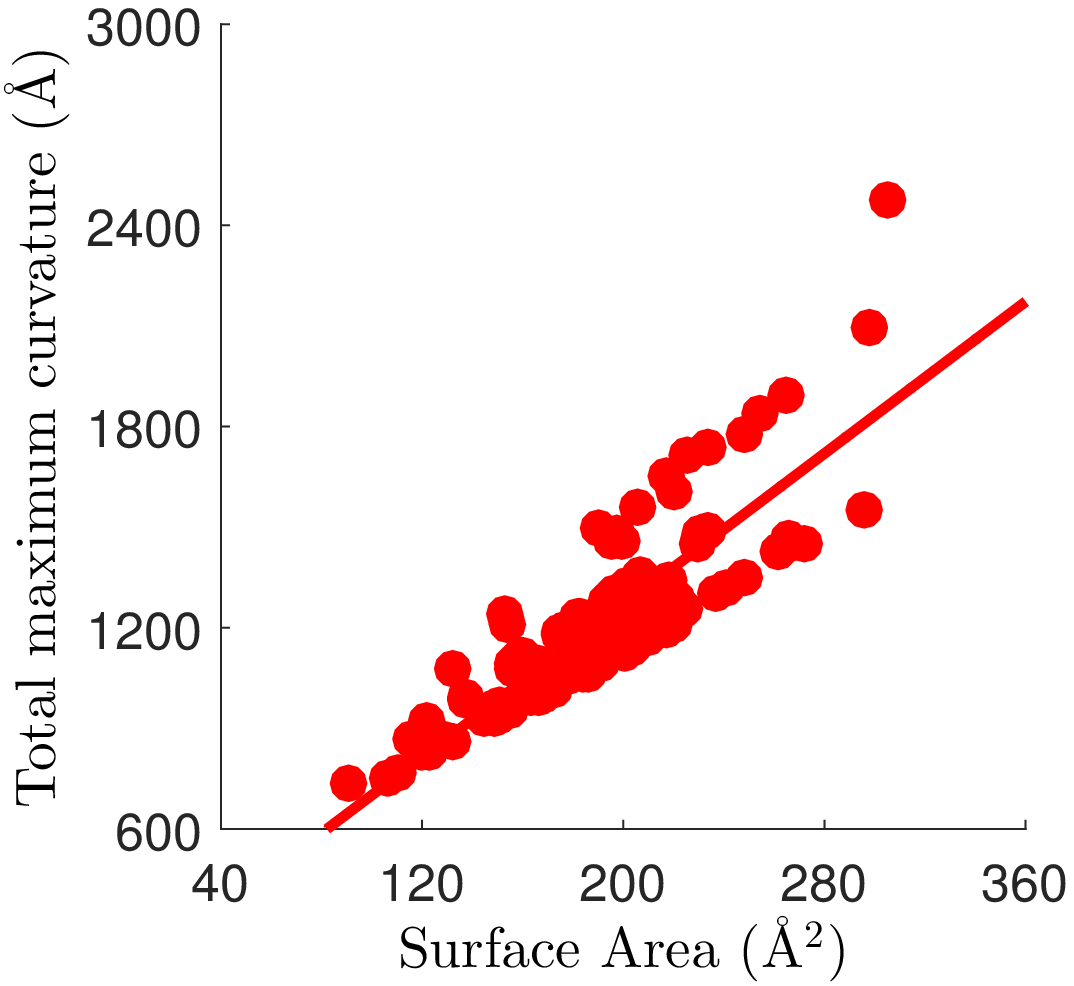}
		\caption{Area versus  curvatures over 127 molecules in all six groups. $R^2$ values of the best fitting lines are 		0.47, 0.22, 0.32 and 0.73, respectively  for  mean, Gaussian, minimum and maximum curvatures.}
		\label{fig.correl_area_meancurv}
	\end{center}
\end{figure}

\begin{figure}[!ht]
	\begin{center}
		\includegraphics[width=0.30\columnwidth]{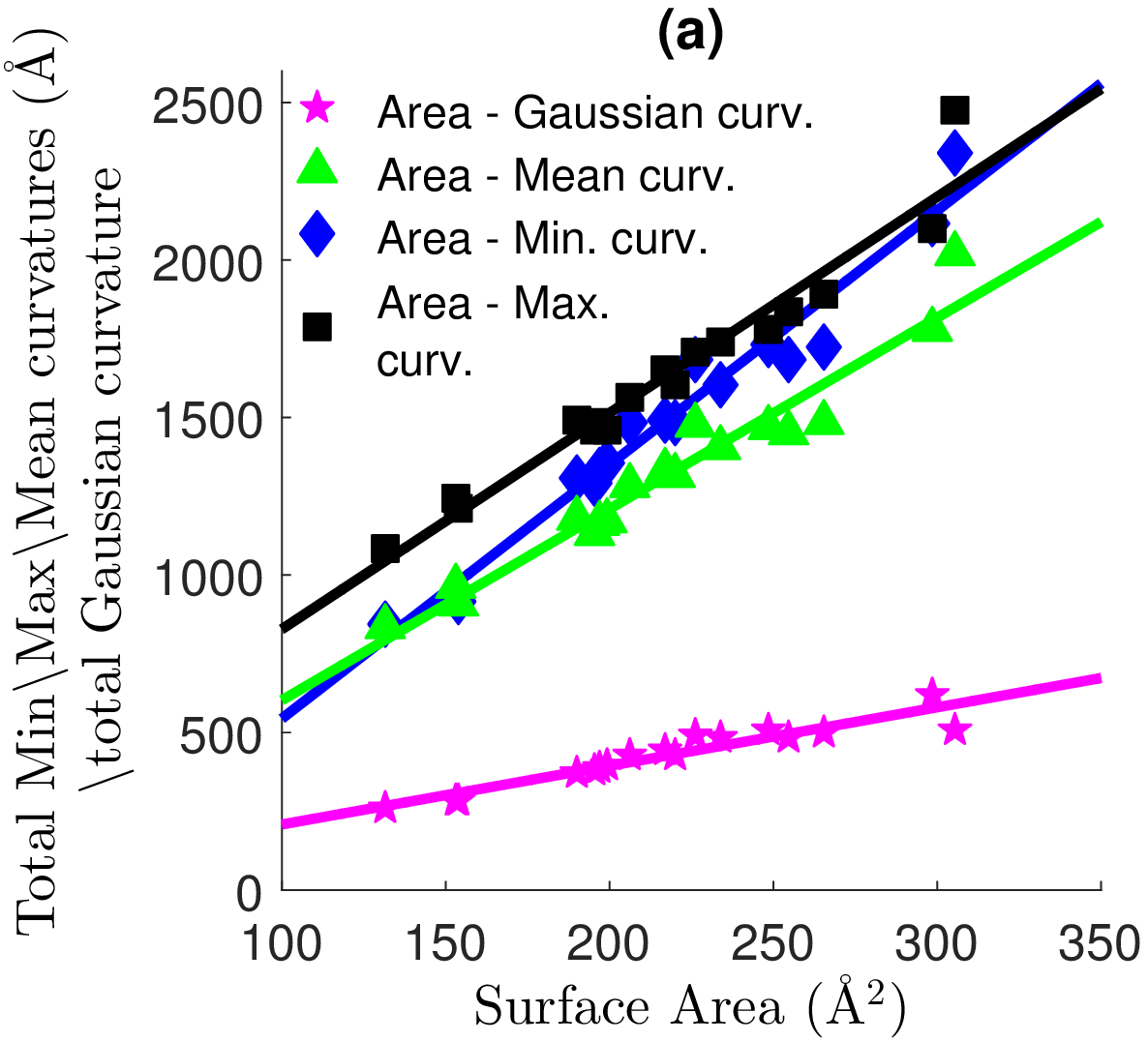}\quad
		\includegraphics[width=0.30\columnwidth]{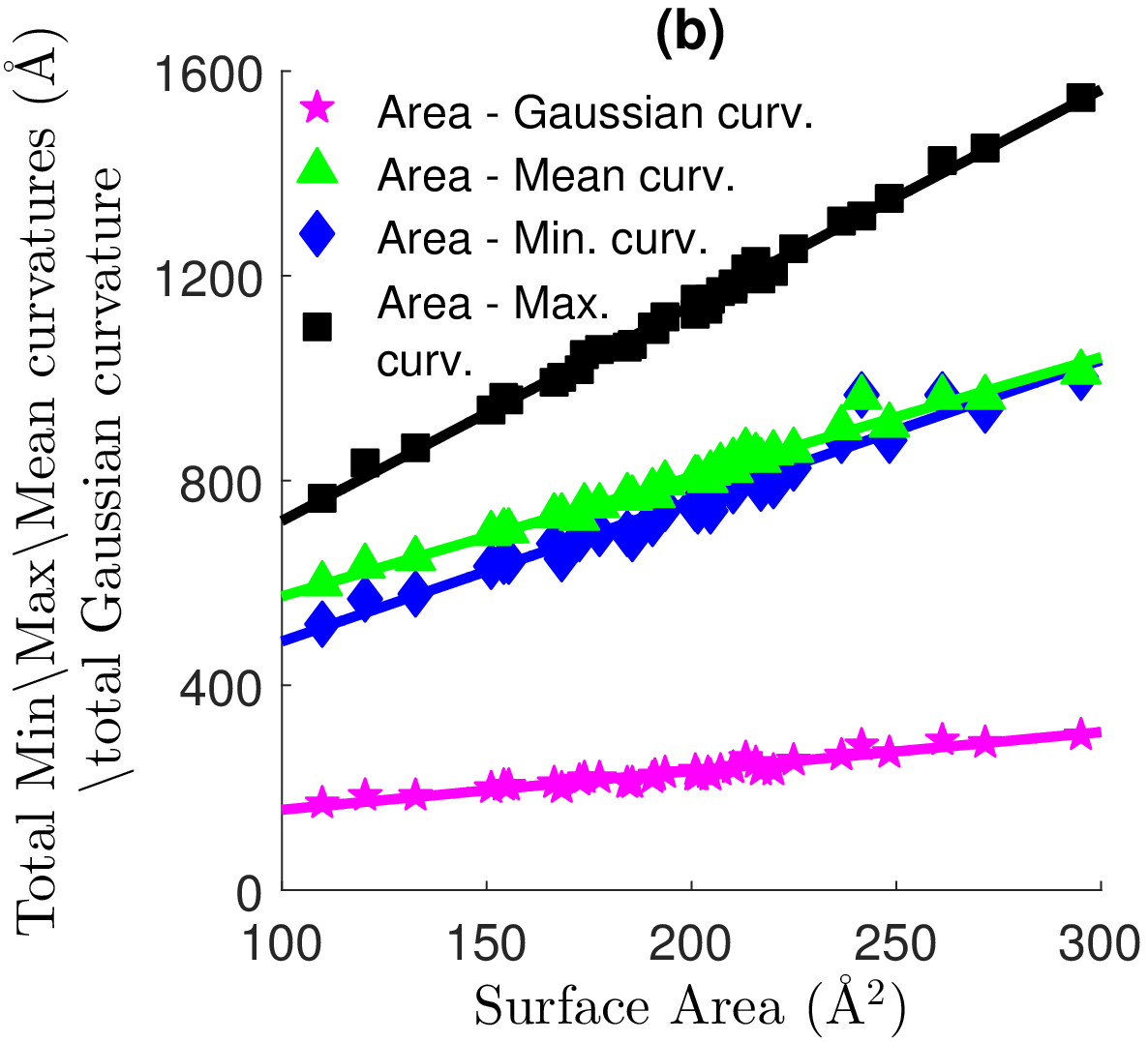}\quad
		\includegraphics[width=0.30\columnwidth]{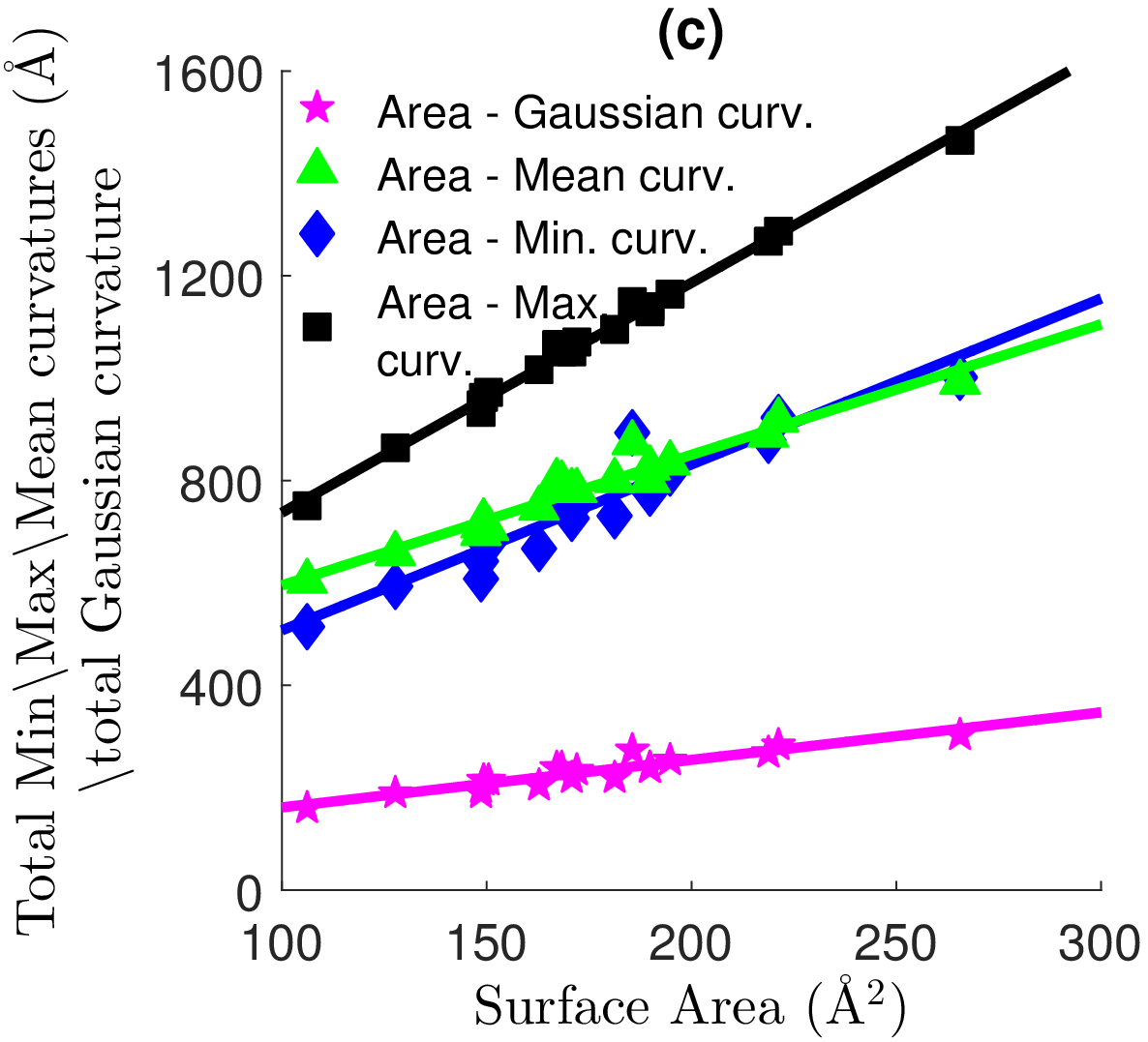}\\
		\vspace*{0.5cm}
		\includegraphics[width=0.30\columnwidth]{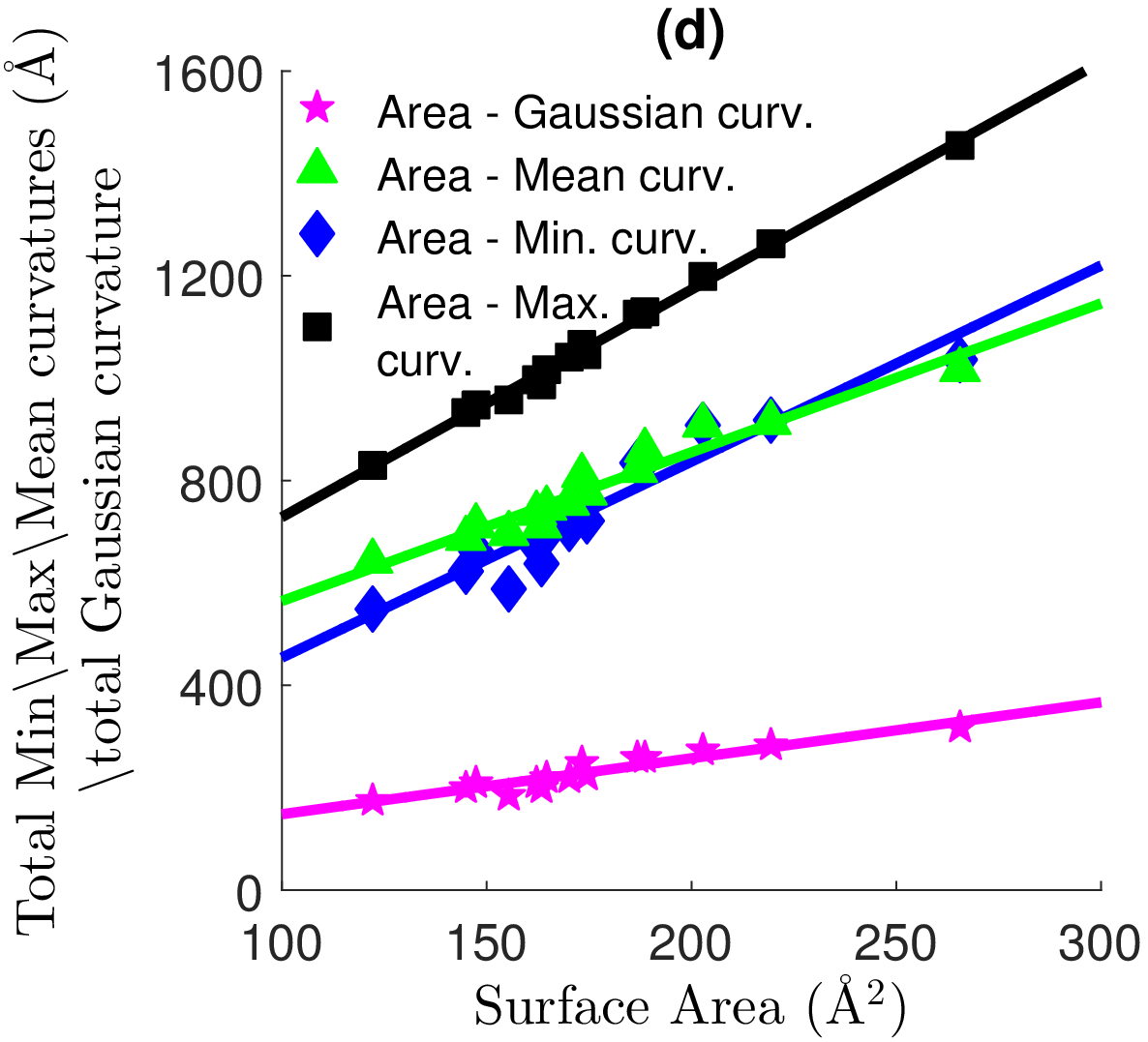}\quad
		\includegraphics[width=0.30\columnwidth]{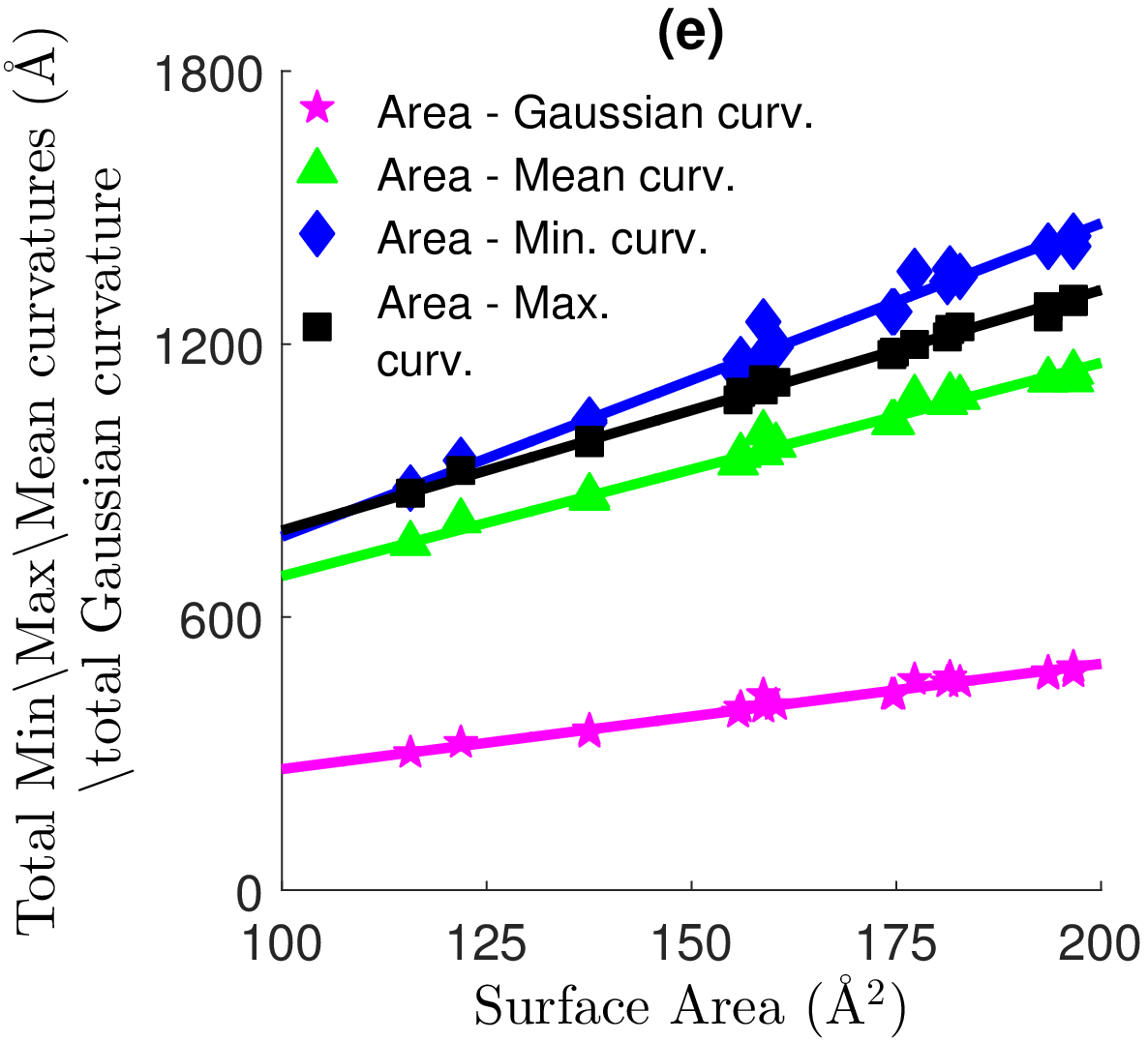}\quad
		\includegraphics[width=0.30\columnwidth]{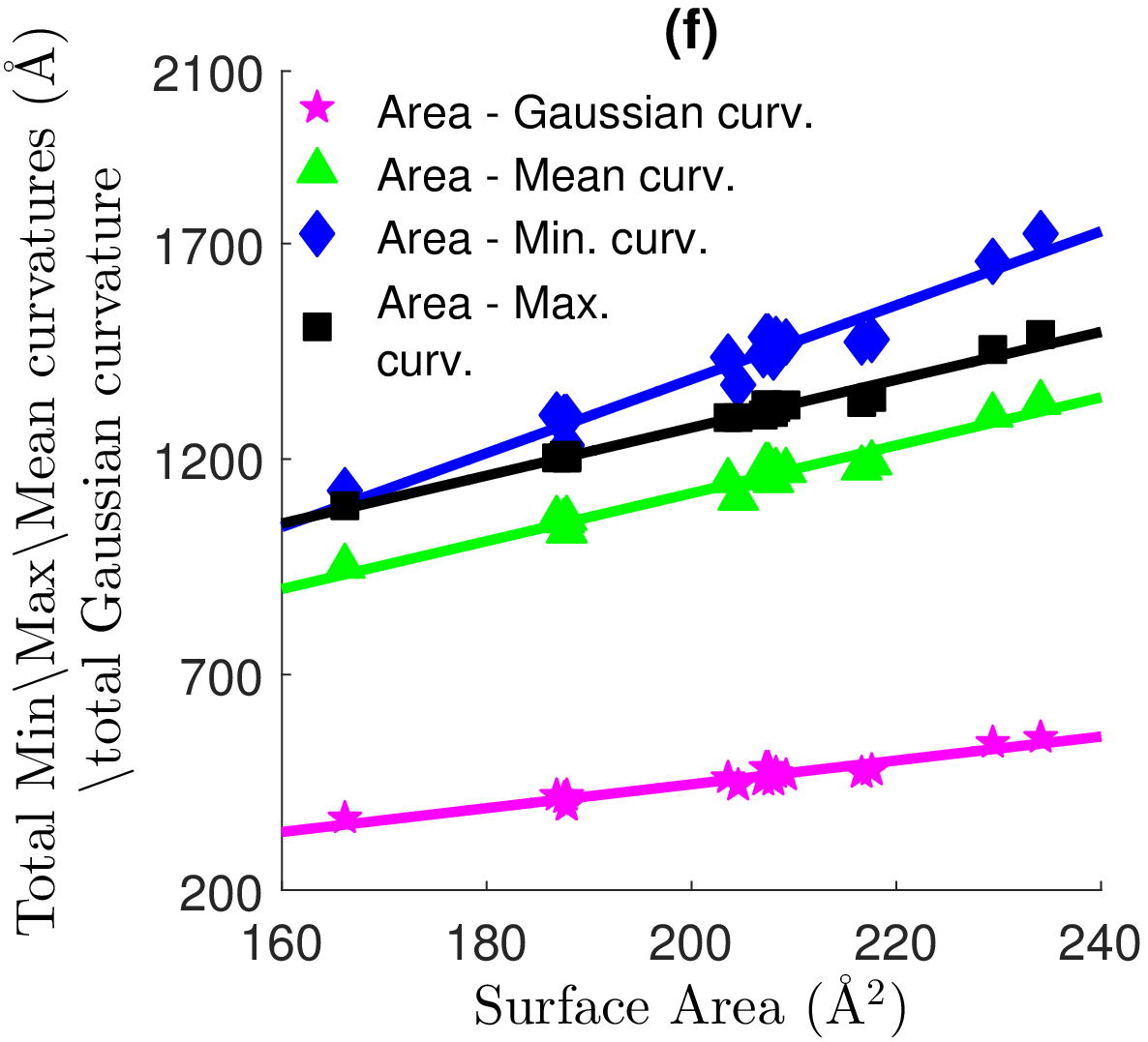}
		\caption{Area versus  minimum, maximum, mean, and Gaussian curvatures. Blue diamond : area versus   minimum curvature, black square: area versus   maximum curvature, green triangle: area versus   mean curvature, pink star: area versus  Gaussian curvature. Six groups are labeled as: (a) SAMPL0 set, (b) alkane set, (c) alkene set, (d) ether set, (e) alcohol set, and (f) phenol set.}
		\label{fig.correl_area_curv}
	\end{center}
\end{figure}

\paragraph{Correlation between areas and volumes}

Figure  \ref{fig.correl_area_volume} shows the correlation between surface areas and surface enclosed volumes for 127 molecules studied in  this work. Apparently, their surface areas and surface enclosed volumes are highly correlated to each other.  The best fitting line and $R^2$   found in this numerical  experiment are, respectively, $y=1.55x-66.51$ and $0.99$. A similar correlation was reported in the literature \cite{David:2019JCTCsolvation}. Therefore, it is computationally inefficient to simultaneously include both area and volume components in a solvation model. However, physically, it is perfectly fine to have both area and volume in a solvation model as surface area represents the energy induced by the surface tension, whereas surface enclosed volume describes the work required to create a cavity in the solvent for a solute molecule. Mathematically, the correlation between  surface areas and volumes of a group of solute molecules can be due to their similarity in their sphericity measurements \cite{KLXia:2015a}.  Therefore, the surface areas and volumes of lipid bilayer sheets will not be correlated with those of micelles or liposomes.

\begin{table}[!ht]
	\centering
	\caption{$R^2$ values and best fitting lines between area and curvature measurements.}
	\label{tab.correl_area_curv}
	\resizebox{\linewidth}{!}{%
		\begin{tabular}{llllllllllll}
			\hline 
			Group  &  \multicolumn{2}{c}{area vs    min. curv.} & & \multicolumn{2}{c}{area vs   max. curv.} & &\multicolumn{2}{c}{area vs   mean curv.} & &\multicolumn{2}{c}{area vs   Gaussian curv.} \\
			\cline{2-3} \cline{5-6} \cline{8-9} \cline{11-12}
			& \multicolumn{1}{c}{fitting line} & \multicolumn{1}{c}{$R^2$} &&  \multicolumn{1}{c}{fitting line} & \multicolumn{1}{c}{$R^2$} &&  \multicolumn{1}{c}{fitting line} & \multicolumn{1}{c}{$R^2$} &&  \multicolumn{1}{c}{fitting line} & \multicolumn{1}{c}{$R^2$} \\
			\hline		   
			SAMPL0 &$ y=8.07x -262.51 $ &      0.96   && $y=6.86x+  141.72 $     &  0.95  &&  $y=6.08x -5.05    $  & 0.95      && $y= 1.86x+ 22.05   $ & 0.90\\
			Alkane &$y=2.75x+  210.87  $ &  0.95       && $y= 4.21x+ 299.83 $     &  0.99  &&  $y=2.34x+  340.21 $  &     0.98  && $y= 0.76x+ 80.84   $ & 0.93\\
			Alkene &$y=3.24x+  183.15 $ &      0.90    && $y=4.49x+  288.34 $     &  0.99  &&  $y= 2.55x+  340.27$  &     0.95  && $y= 0.93x+   68.51  $ &     0.87\\
			Ether  &$y=3.83x+70.92    $ &  0.91       && $y= 4.45x+  283.94$     &  0.99  &&  $y=2.91x+  273.88 $  &     0.94  && $y= 1.09x+   38.78 $ &     0.91\\
			Alcohol&$y=6.89x+87.63    $ &  0.99       && $y=5.29x+261.34   $     &  1.00  &&  $y=4.69x+221.01   $  &     0.99  && $y=2.32x+34.15     $ &     0.99 \\
			Phenol &$y=8.58x-330.11   $ &  0.94       && $y=5.56x+161.15   $     &  0.98  &&  $y=5.56x+9.16     $  &     0.95  && $y=2.77x-108.17    $ &     0.93\\
			\hline
		\end{tabular}	}
	\end{table}

	\begin{figure}[!ht]
		\begin{center}
			\includegraphics[width=0.30\columnwidth]{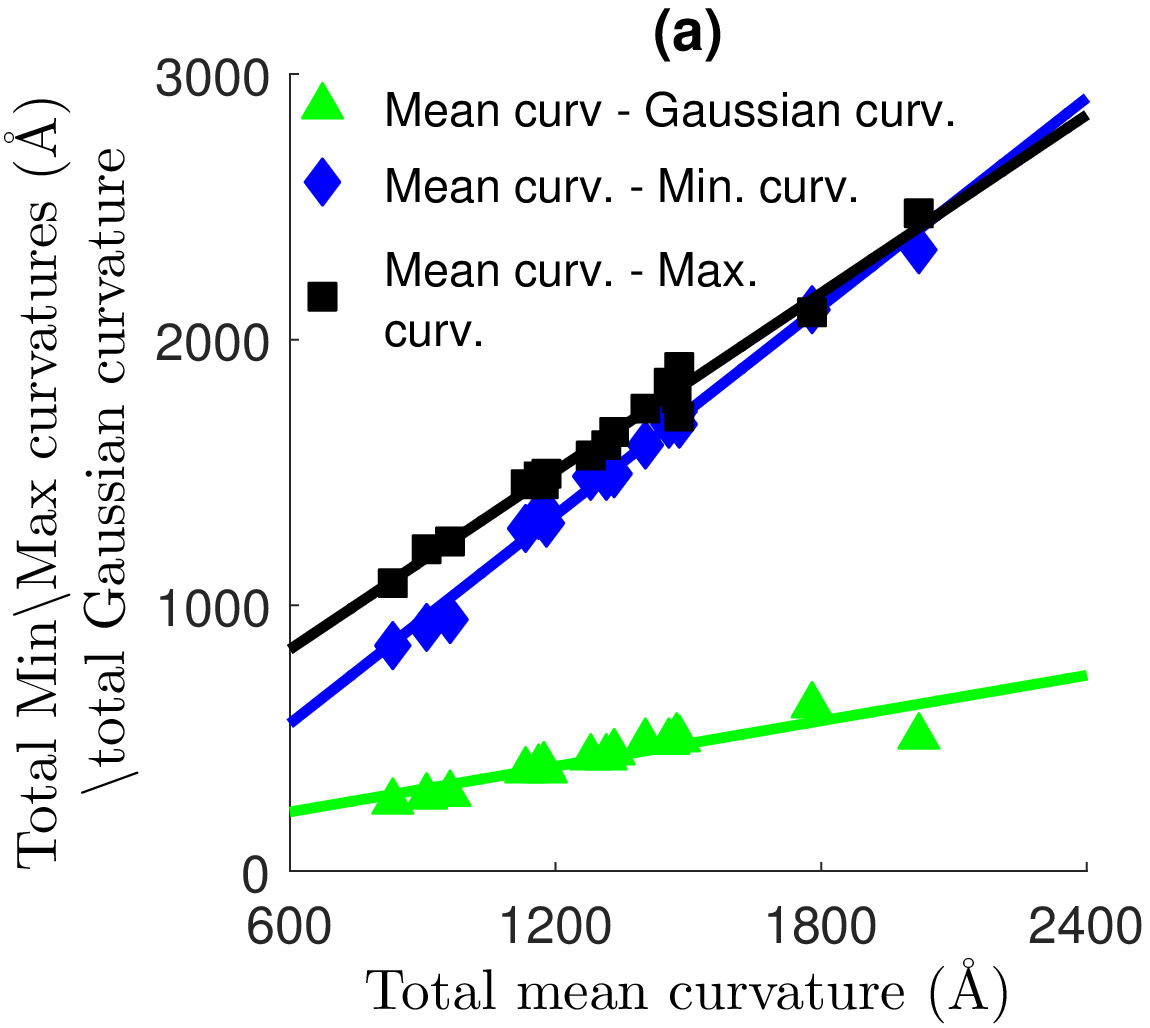}\quad
			\includegraphics[width=0.30\columnwidth]{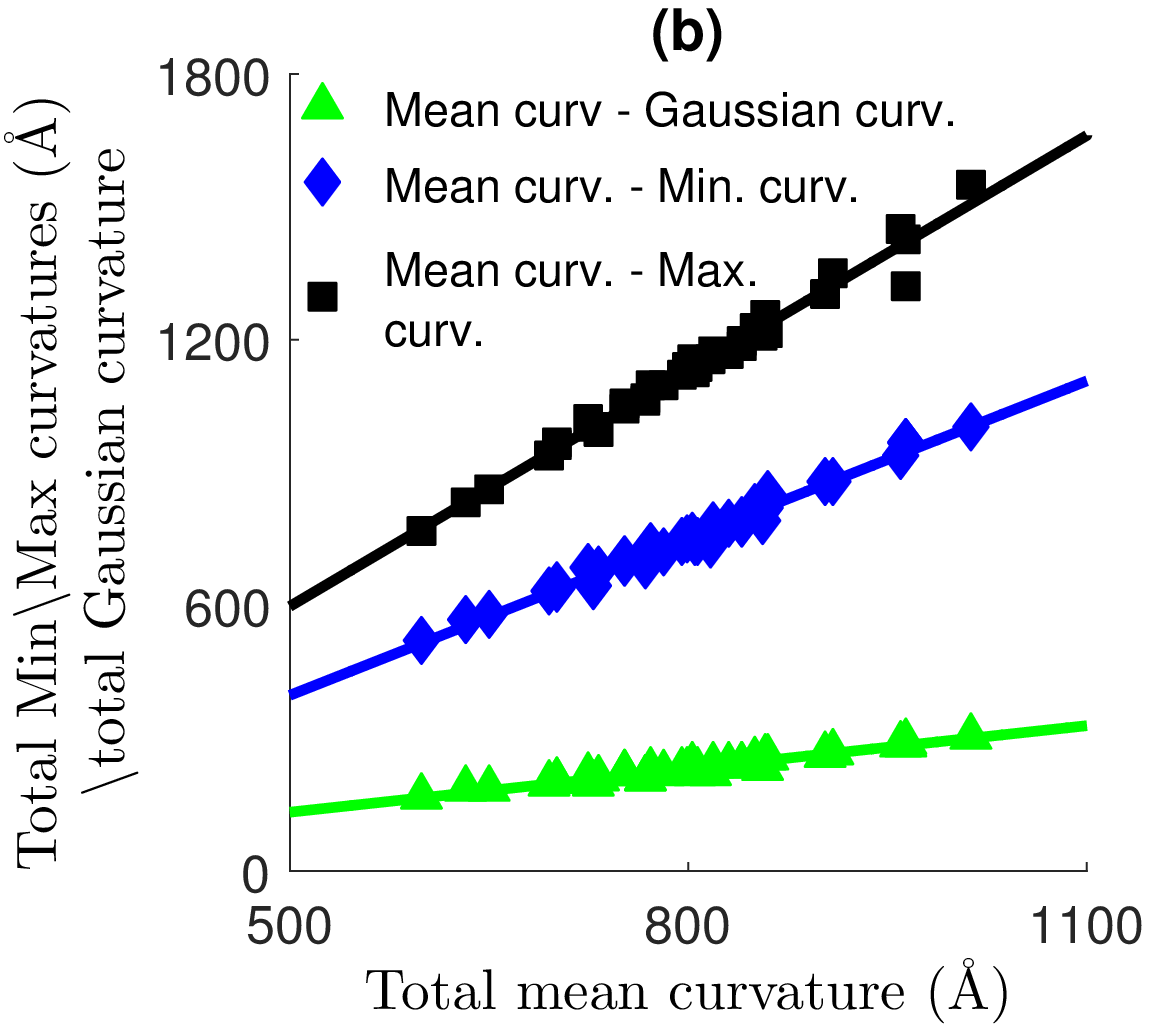}\quad
			\includegraphics[width=0.30\columnwidth]{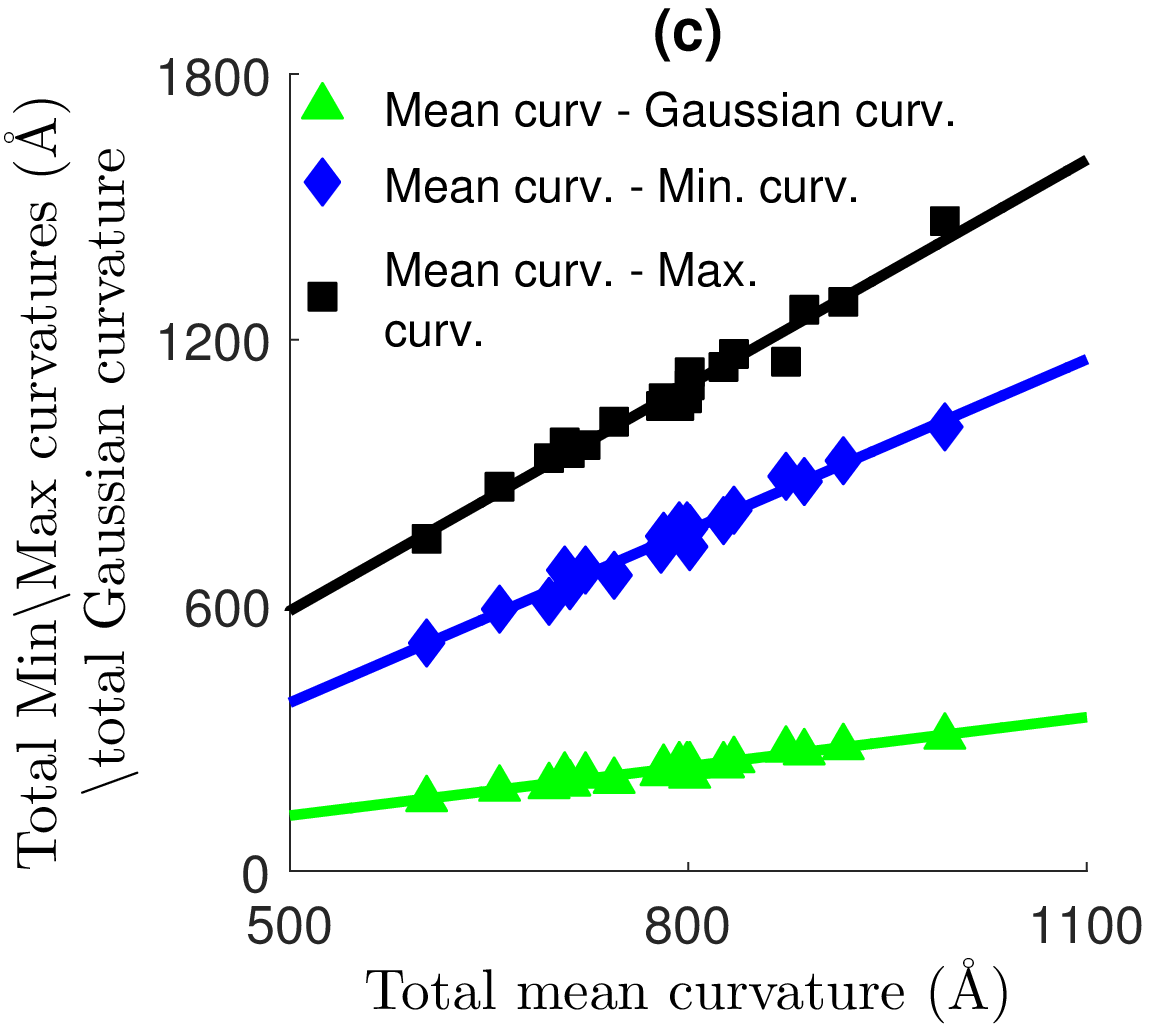}\\
			\vspace*{0.5cm}
			\includegraphics[width=0.30\columnwidth]{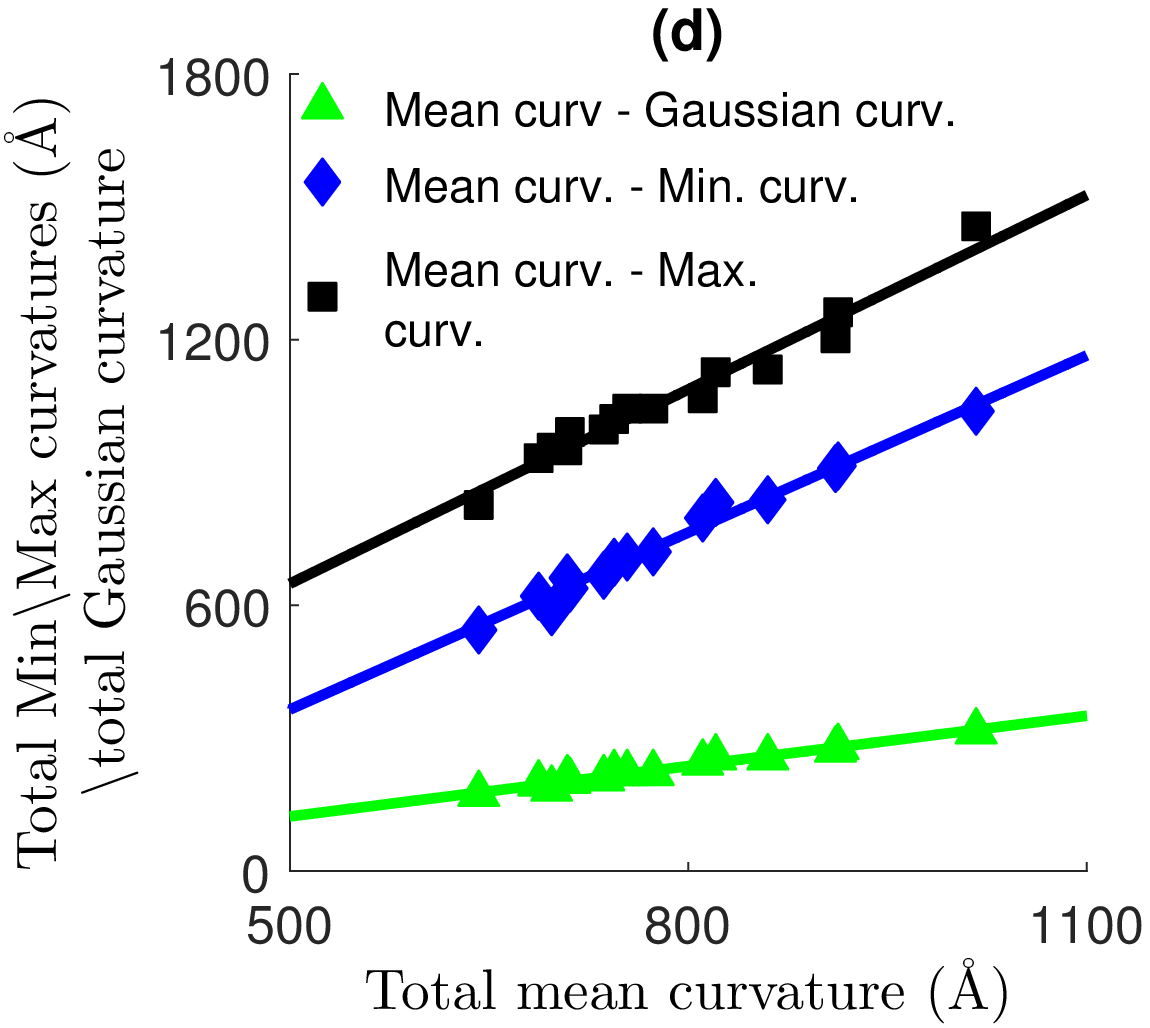}\quad
			\includegraphics[width=0.30\columnwidth]{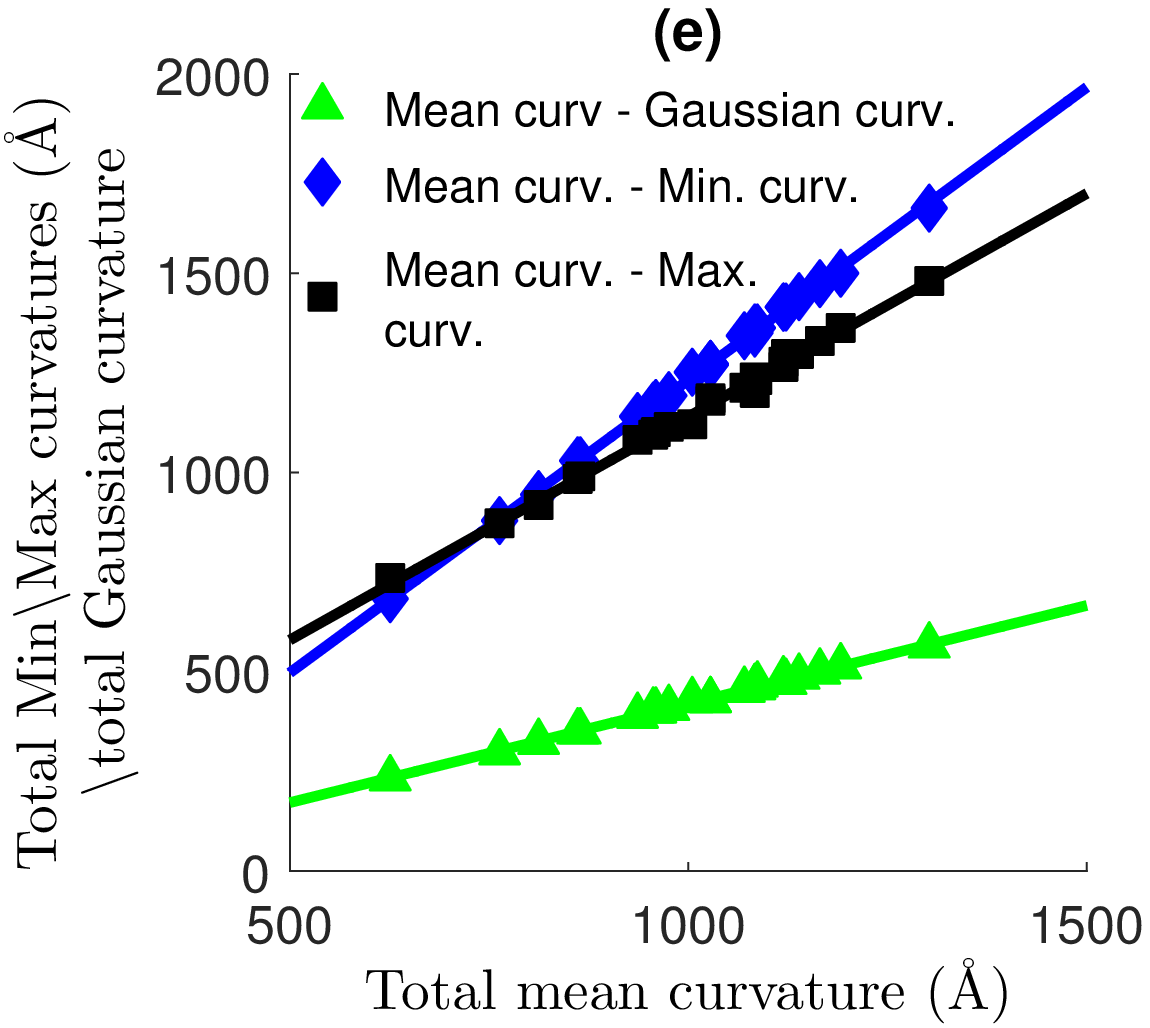}\quad
			\includegraphics[width=0.30\columnwidth]{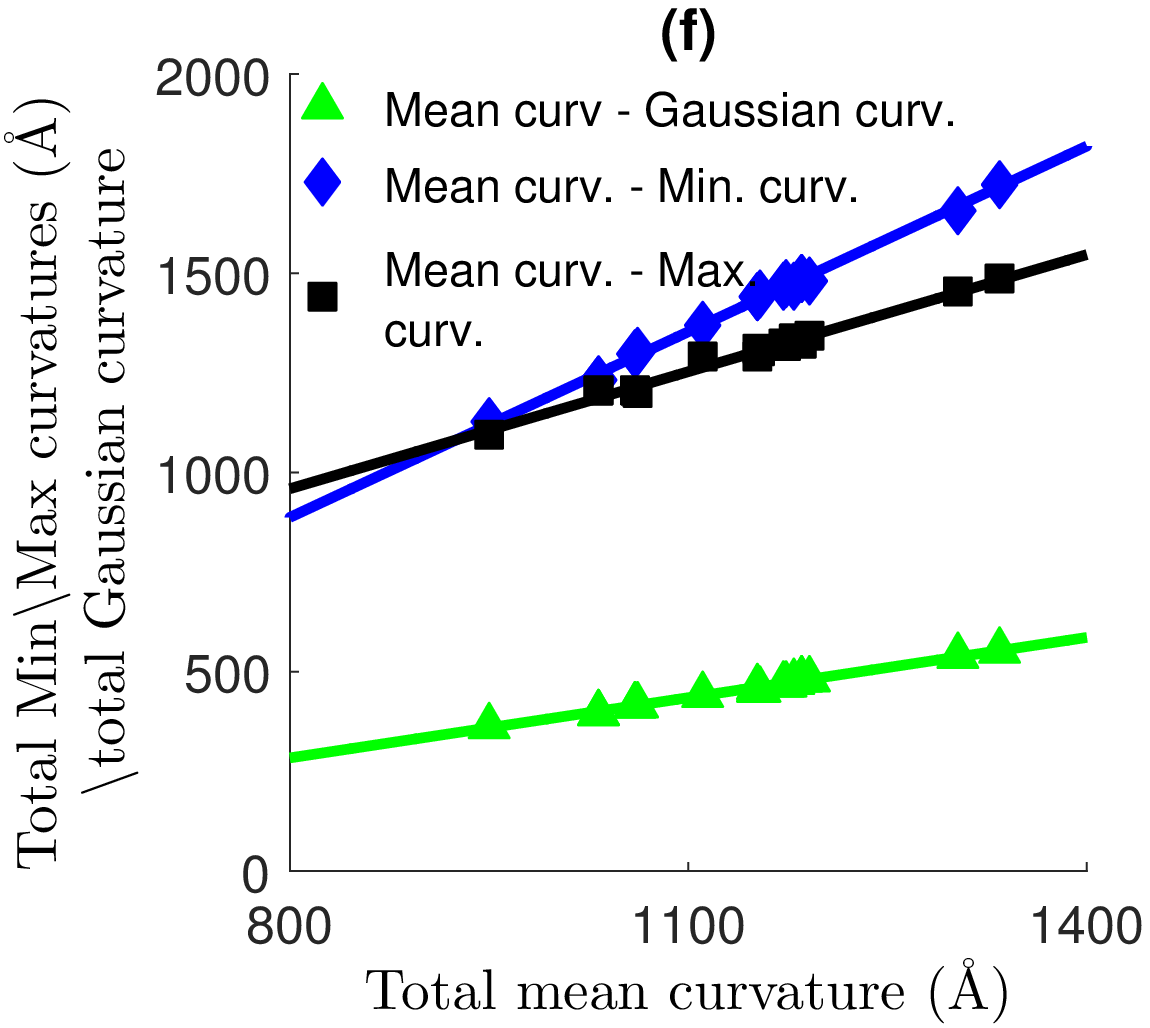}
			\caption{Mean curvature versus   minimum, maximum, and Gaussian curvatures. Green triangle:   mean curvature versus   Gaussian curvature, blue diamond:   mean curvature versus   minimum curvature, black square:  mean curvature versus   maximum curvature. Six groups are labeled as: (a) SAMPL0set, (b) alkane set, (c) alkene set, (d) ether set, (e) alcohol set, and (f) phenol set.}
			\label{fig.correl_mean_curv}
		\end{center}
	\end{figure} 
	
	\begin{table}[!ht]
		\centering
		\caption{$R^2$ values and best fitting lines between   mean curvature and another types of   curvatures.}
		\label{tab.correl_mean_curv}
		\resizebox{\linewidth}{!}{%
			\begin{tabular}{lllllllll}
				\hline 
				Group  &  \multicolumn{2}{c}{  mean curv. vs   min. curv.} & & \multicolumn{2}{c}{  mean curv. vs    max. curv.} & &\multicolumn{2}{c}{  mean curv. vs   Gaussian curv.} \\
				\cline{2-3} \cline{5-6} \cline{8-9}
				& \multicolumn{1}{c}{fitting line} & \multicolumn{1}{c}{$R^2$} &&  \multicolumn{1}{c}{fitting line} & \multicolumn{1}{c}{$R^2$} &&  \multicolumn{1}{c}{fitting line} & \multicolumn{1}{c}{$R^2$} \\
				\hline		   
				SAMPL0 &$y=1.42x  -34.72$&  0.99 && $y=1.16x+  19.71$ & 0.98  && $y=0.54x  -12.48$ & 0.97\\
				Alkane &$y=1.19x  -32.63$&  0.99&&  $y=1.79x  -49.63$ & 0.99  && $y=0.34x   -4.92$ & 0.96\\
				Alkene &$y=1.27x  -40.51$&  0.98&&  $y=1.70x  -42.13$ & 0.98  && $y=0.38x   -8.32$ & 0.96\\
				Ether  &$y=1.33x  -49.84$&  0.99&&  $y=1.52x  -19.49$ & 0.97  && $y=0.40x  -12.01$ & 0.98\\
				Alcohol&$y=1.52x  -19.20$&  1.00&&  $y=1.08x+   5.87$ & 1.00  && $y=0.89x  -13.79$ & 1.00\\
				Phenol &$y=1.57x  -26.77$&  1.00&&  $y=1.03x+  17.22$ & 0.98  && $y=0.87x  -18.57$ & 0.99\\
				\hline
			\end{tabular}	}
		\end{table}
		
		\paragraph{Correlation between areas and curvatures}
		
		We next investigate the correlations between surface areas and four different types of curvatures for 127 molecules. Our results are depicted in Fig. \ref{fig.correl_area_meancurv}.  
		Obviously,  the correlation between surface areas and  maximum curvatures is the highest among curvature counterparts. The 
		$R^2$ value for the best fitting line is 0.73. However, mean curvatures, Gaussian curvatures and minimum curvatures do not relate to surface areas very well.   Their  $R^2$ values for the best fitting lines are 0.47, 0.22 and 0.32, respectively, which are unsatisfactory.
		
		These results are expected because maximum curvatures are mostly rendered from  the convex surfaces of the molecular rigidity surface manifold, whereas minimum curvatures correspond to the  concave  surfaces of the molecular rigidity surface manifold. Topologically, in spirit of Morse-Smale theory, a family of extreme values of minimum curvatures defined at various isosurfaces gives rise to a natural decomposition of molecular rigidity density and leads to ``rigidity complex''.  
		The mean curvature is the average of minimum and maximum curvatures. The Gaussian curvature, as the product of two principle curvatures, correlates  the least to the surface area for 127 molecules studied.  
		Therefore, compared to volumes,  Gaussian   and minimum curvatures are  complementary to surface areas and thus, are more useful for solvation modeling in general.

		However, a careful examination of   Fig. \ref{fig.correl_area_meancurv} reveals certain linear features. To understand the origin of the data alignment in  Fig. \ref{fig.correl_area_meancurv}, we analyze the correlations between  surface areas and curvatures in six test sets.  Figure \ref{fig.correl_area_curv} depicts these correlations. Obviously, there are good correlations in each test set.  The best fitting lines and $R^2$ values of the corresponding date are reported in Table \ref{tab.correl_area_curv}. These data further indicate  that surface area  and  curvature quantities  in each test set are well correlated; specifically, $R^2$  values of them are always larger than $0.89$. By averaging over six groups, the   maximum curvature has the highest correlation with surface area, following by mean curvature, minimum curvature and Gaussian curvature. Surprisingly, for mean, Gaussian   and minimum curvatures, such well correlations only occur in  individual test sets. 
		
		Moreover, the slopes of fitting lines in Table \ref{tab.correl_area_curv} indicates that the curvatures and areas in alkane, alkene and ether sets are well correlated. A possible  reason for this correlation is that structures of the molecules in these three groups are quite similar to each other.

		\paragraph{Correlation between different curvatures}
		
		Additionally, we are  interested in finding the correlations between different curvatures. Such a finding enables us to determine how many curvature terms in an efficient solvation model.  Figure \ref{fig.correl_mean_curv} depicts the correlation data between   mean curvature and other types of  curvatures for each group. As expected, different types of curvature are correlated to each other extremely well for each group. Table \ref{tab.correl_mean_curv} provides the best fitting lines and $R^2$ values for such correlations, and we can see that $R^2$ for any case is always higher than $0.95$. Based on this correlation analysis, it is clear that different curvatures will have the same modeling effect in solvation analysis and thus at most one type of curvature term is needed in an efficient solvation model.  The correlations among different curvatures for all 127 molecules are illustrated  in Fig. S1 in Supporting Information. 
		
		
		
\subsection{The influence of surface area, volume, curvatures and Lennard-Jones potential on the accuracy of solvation free energy prediction}
		
		\begin{table}[!hb]
 
				\caption{The solvation free energy prediction for the SAMPL0 set with different models. Energy is in the unit of kcal/mol.}
			\label{tab.sampl0_eng}
			\resizebox{\linewidth}{!}{%
				\begin{tabular}{rrrrrrrrrrrrrrrrrrr}
					\hline 
					& & M01 & M02 & M03& M04& M05& M06& M07& M08& M09& M10& M11& M12& M13& M14& M15& M16& M17\\
					\hline
					& \multicolumn{1}{l}{$\Delta G^{\text{Exp}}$\cite{Nicholls:2008solvation}} &-8.84	& -2.38	 & -1.93  &	1.07  &	-11.01&	-9.76&	-4.23	&-4.97&	-3.28&	-5.05&	-6.00&	-2.93&	-6.34&	-3.54&	-1.43&	-4.08&	-9.81\\
					& \multicolumn{1}{l}{$\Delta G^{\text{p}}$} &-5.27	& -2.10	 & -2.17  &	-1.45	&-4.43	&-3.82	&-1.52	&-3.78	&-0.99	&-1.98	&-3.54	&-1.37	&-3.45	&-0.97	&-1.14	&-3.43	&-4.93\\
					\hline							  
					\multirow{3}{*}{$\mathbf{H}$} & \multicolumn{1}{l}{$\Delta G^{\mathbf{H}}$}&-2.79	&-1.83	&-1.78&	-3.17&	-2.33	&-2.29	&-2.01	&-2.32	&-2.09	&-1.43	&-2.31	&-1.51&	-2.07	&-2.20	&-1.85	&-1.85&	-1.31\\
					&  \multicolumn{1}{l}{$\Delta G$}&-8.06	&-3.93	&-3.95&	-4.62&	-6.76	&-6.10	&-3.54	&-6.10	&-3.08	&-3.41	&-5.85	&-2.89&	-5.52	&-3.18	&-2.99	&-5.27&	-6.24\\
					&  \multicolumn{1}{l}{Error}&-0.78	&1.55	&2.02&	5.69	&-4.25	&-3.66	&-0.69	&1.13	&-0.20	&-1.64	&-0.15	&-0.04&	-0.82	&-0.36	&1.56	&1.19	&-3.57\\
					\cline{2-19}
					&  \multicolumn{1}{l}{RMSE} &\multicolumn{17}{c}{2.34}\\
					\hline					
					\multirow{3}{*}{$\mathbf{A}$} & \multicolumn{1}{l}{$\Delta G^{\mathbf{A}}$}	&-2.94	&-1.94&	-1.92	&-3.01	&-2.61	&-2.50	&-2.03&	-2.22&-2.14	&-1.52	&-2.45&	-1.51	&-2.17	&-2.31&-1.88	&-1.96	&-1.30\\
					& \multicolumn{1}{l}{$\Delta G$}&-8.21	&-4.04&	-4.09	&-4.45	&-7.04	&-6.32	&-3.55&	-6.00&-3.13	&-3.50	&-5.99&	-2.88	&-5.62	&-3.28	&-3.02	&-5.39	&-6.23\\
					&  \multicolumn{1}{l}{Error}&-0.63	&1.66	&2.16	&5.52	&-3.97&	-3.44&	-0.68&	1.03	&-0.15	&-1.55&	-0.01	&-0.05&	-0.72	&-0.26	&1.59	&1.31	&-3.58\\
					\cline{2-19}  
					&  \multicolumn{1}{l}{RMSE} &\multicolumn{17}{c}{2.27}\\
					\hline
					\multirow{3}{*}{$\mathbf{L}$} & \multicolumn{1}{l}{$\Delta G^{\mathbf{L}}$}&-3.37	&-0.28	&-1.79	&2.52	&-4.29	&-4.21	&-2.36	&-2.49&	-2.99&	-1.96	&-2.89	&-1.98	&-2.57&	-3.13	&-0.29	&-1.76	&-6.03\\
					& \multicolumn{1}{l}{$\Delta G$} &-8.64	&-2.38	&-3.96	&1.07	&-8.72	&-8.02	&-3.88	&-6.27&	-3.98&	-3.94	&-6.43	&-3.36	&-6.02&	-4.10	&-1.43	&-5.19	&-10.96\\
					&  \multicolumn{1}{l}{Error}&-0.20	&0.00	&2.03	&0.00	&-2.29	&-1.74	&-0.35	&1.30&	0.70&	-1.11	&0.43	&0.43	&-0.32&	0.56	&0.00	&1.11	&1.15\\
					\cline{2-19}  
					&  \multicolumn{1}{l}{RMSE} &\multicolumn{17}{c}{1.07}\\
					\hline 
					\multicolumn{1}{c}{\multirow{3}{*}{\makecell{$\mathbf{A}$\\$\mathbf{H}$}}}& \multicolumn{1}{l}{$\Delta G^{\mathbf{A}}$} &-40.93&-27.04&	-26.78	&-41.87	&-36.39	&-34.89&-28.24	&-30.98&-29.79&	-21.16&	-34.10&	-21.03	&-30.23	&-32.14&-26.13	&-27.36	&-18.10\\
					& \multicolumn{1}{l}{$\Delta G^{\mathbf{H}}$}&37.41	&24.46&	23.83	&42.47	&31.18	&30.61	&26.95	&31.13	&28.01&	19.12&	30.96&	20.27	&27.66	&29.52	&24.79	&24.74	&17.55\\
					& \multicolumn{1}{l}{$\Delta G$}&-8.79	&-4.68&	-5.11	&-0.85	&-9.64	&-8.10	&-2.82	&-3.64	&-2.77&	-4.02&	-6.68&	-2.13	&-6.02	&-3.58	&-2.47	&-6.04	&-5.48\\
					&  \multicolumn{1}{l}{Error}&-0.05	&2.30&	3.18	&1.92	&-1.37	&-1.66	&-1.41	&-1.33	&-0.51&	-1.03&	0.68	&-0.80	&-0.32	&0.04	&1.04	&1.96	&-4.33\\
					\cline{2-19}
					&  \multicolumn{1}{l}{RMSE} &\multicolumn{17}{c}{1.78}\\
					\hline 
					\multicolumn{1}{c}{\multirow{3}{*}{\makecell{$\mathbf{H}$\\$\mathbf{L}$}}}& \multicolumn{1}{l}{$\Delta G^{\mathbf{H}}$} &27.06	&17.69	&17.23	&30.71	&22.55	&22.14	&19.49	&22.51	&20.26	&13.83	&22.39&	14.66	&20.01	&21.35	&17.93&	17.89	&12.69\\
					& \multicolumn{1}{l}{$\Delta G^{\mathbf{L}}$}&-31.17	&-17.97	&-17.47	&-28.20	&-28.74	&-27.41	&-22.11	&-22.81&-23.02	&-16.59&-25.41&	-15.62&	-23.01	&-24.09&-18.22&	-18.77&	-17.87\\
					& \multicolumn{1}{l}{$\Delta G$} &-9.38	&-2.38	&-2.40	&1.07	&-10.61	&-9.09	&-4.15	&-4.07	&-3.75	&-4.74	&-6.55&	-2.34	&-6.45	&-3.71	&-1.43&	-4.31	&-10.11\\
					&  \multicolumn{1}{l}{Error}&0.54	&0.00	&0.47	&0.00	&-0.40	&-0.67	&-0.08	&-0.90	&0.47	&-0.31	&0.55&	-0.59	&0.11	&0.17	&0.00&	0.23	&0.30\\
					\cline{2-19}
					&  \multicolumn{1}{l}{RMSE} &\multicolumn{17}{c}{0.43}\\
					\hline 
					\multicolumn{1}{c}{\multirow{3}{*}{\makecell{$\mathbf{A}$\\$\mathbf{H}$\\$\mathbf{L}$}}}& \multicolumn{1}{l}{$\Delta G^{\mathbf{A}}$} &25.16	&16.62	&16.46	&25.74	&22.37	&21.45	&17.36	&19.05	&18.31	&13.01	&20.96	&12.93	&18.58&	19.75	&16.06	&16.82	&11.13\\
					& \multicolumn{1}{l}{$\Delta G^{\mathbf{H}}$} &15.70	&10.26	&10.00	&17.82	&13.08	&12.84	&11.31	&13.06	&11.75	&8.02	&12.99	&8.50	&11.61&	12.39	&10.40	&10.38	&7.36\\
					& \multicolumn{1}{l}{$\Delta G^{\mathbf{L}}$}&-44.94	&-27.17	&-26.35	&-41.04	&-41.61	&-39.87	&-31.35	&-32.88	&-33.15	&-23.93	&-36.59	&-22.18&-33.03&	-34.67&	-26.75	&-28.12	&-23.60\\
					& \multicolumn{1}{l}{$\Delta G$} &-9.35	&-2.38	&-2.06	&1.07	&-10.58	&-9.40	&-4.21	&-4.55	&-4.08	&-4.88	&-6.17	&-2.12	&-6.29&	-3.50	&-1.43	&-4.35	&-10.04\\
					&  \multicolumn{1}{l}{Error}&0.51	&0.00	&0.13	&0.00	&-0.43	&-0.36	&-0.02	&-0.42	&0.80	&-0.17	&0.17	&-0.81	&-0.05&	-0.04	&0.00	&0.27	&0.23\\
					\cline{2-19}
					&  \multicolumn{1}{l}{RMSE} &\multicolumn{17}{c}{0.36}\\
					\hline
					\multicolumn{1}{c}{\multirow{3}{*}{\makecell{$\mathbf{A}$\\$\mathbf{V}$\\$\mathbf{H}$\\$\mathbf{L}$}}}& \multicolumn{1}{l}{$\Delta G^{\mathbf{A}}$}&21.86	&14.44	&14.30	&22.36	&19.44	&18.63&	15.08	&16.55	&15.91	&11.30	&18.22	&11.23	&16.15	&17.16	&13.95	&14.61	&9.67\\
					& \multicolumn{1}{l}{$\Delta G^{\mathbf{V}}$}&4.46	&2.69	&2.67	&5.07	&3.90	&3.73	&2.69	&3.12	&2.95	&1.95	&3.61	&1.87	&3.16	&3.13	&2.54	&2.74	&1.54\\
					& \multicolumn{1}{l}{$\Delta G^{\mathbf{H}}$}&17.68	&11.56	&11.26	&20.07	&14.73	&14.46&	12.73	&14.71	&13.24	&9.04	&14.63	&9.58	&13.07	&13.95	&11.71	&11.69	&8.29\\
					& \multicolumn{1}{l}{$\Delta G^{\mathbf{L}}$}&-47.99	&-28.97	&-28.08	&-44.98	&-44.22	&-42.33&-33.20	&-35.10	&-35.15	&-25.47	&-39.00	&-23.55	&-35.24	&-36.76&-28.50	&-30.11&-24.63\\
					& \multicolumn{1}{l}{$\Delta G$} &-9.26	&-2.38	&-2.02	&1.07	&-10.58	&-9.32&	-4.21	&-4.49	&-4.04	&-5.16	&-6.08	&-2.24	&-6.31	&-3.49	&-1.43	&-4.50	&-10.06\\
					&  \multicolumn{1}{l}{Error}&0.42	&0.00	&0.09	&0.00	&-0.43	&-0.44&	-0.02	&-0.48	&0.76	&0.11	&0.08	&-0.69	&-0.03	&-0.05	&0.00	&0.42	&0.25\\
					\cline{2-19}
					&  \multicolumn{1}{l}{RMSE} &\multicolumn{17}{c}{0.35}\\
					\hline
				\end{tabular}}		
				{\small   
					M01:    Glycerol triacetate; 		 
					M02: Benzyl bromide; 		 
					M03:  Benzyl chloride; 	
					M04:   m-bis (trifluoromethyl) benzene; 
					M05:  N,N-dimethyl-p-methoxybenz; 				
					M06:  N,N-4-trimethylbenzamide;	 
					M07:  bis-2-chloroethyl ether; 	 
					M08:  1,1-diacetoxyethane; 		 
					M09:  1,1-diethoxyethane; 		 
					M10:   1,4-dioxane; 				 
					M11:  Diethyl propanedioate; 	 
					M12:   Dimethoxymethane; 			 
					M13:  Ethylene glycol diacetate;
					M14:  1,2-diethoxyethane;		 
					M15:   Diethyl sulfide; 			 
					M16:   Phenyl formate; 			 
					and 
					M17: Imidazole. }				 
			\end{table}

			To examine the impact of area, volume, curvature and Lennard-Jones potential in the solvation prediction, we firstly explore seven different models including $\mathbf{H}$, $\mathbf{A}$, $\mathbf{L}$, $\mathbf{AH}$, $\mathbf{HL}$, $\mathbf{AHL}$, and $\mathbf{AVHL}$ to  predict the solvation free energy for SAMPL0 test set. For the sake of simplicity, we use short notations to represent 17 molecules in SAMPL0 test set, and their full names are given in the caption of Table  \ref{tab.sampl0_eng}. Judging by RMS errors  evaluated between the experimental and predicted solvation free energies, Table \ref{tab.sampl0_eng} reveals that Lennard-Jones potential plays an important role in the accuracy of the solvation free energy prediction. If we only consider this term in the nonpolar calculation, i.e., model $\mathbf{L}$, the RMS error for this case is as low as $1.07$ kcal/mol, which is a very reasonable result in comparison to those  reported in the literature, such as $0.60$ kcal/mol in \cite{BaoWang:2015a}, and $1.71\pm0.05$ kcal/mol in \cite{Nicholls:2008solvation}. On the other hand, if the Lennard-Jones potential is absent in nonpolar calculations, the solvation free energy prediction   performs poorly for SAMPL0. To be specific, the RMS errors for models $\mathbf{H}$, $\mathbf{A}$, and $\mathbf{AH}$ listed in Table \ref{tab.sampl0_eng}  are all over $1.75$ kcal/mol. As the previous analysis in Section \ref{sec.correl}, mean curvature and area are well correlated; therefore, the RMS errors for models $\mathbf{H}$ and $\mathbf{A}$  are very similar and are, respectively, $2.34$ and $2.27$. Even the combination of them in model $\mathbf{AH}$ does not improve the solvation prediction very much, and its RMS error is found to be $1.78$. Due to correlations, models involving only different types of curvatures and volume will have the similar results (data not shown). On the other hand, the mixture of Lennard-Jones potential and other quantities can significantly improve the solvation prediction accuracy. To be specific, Table \ref{tab.sampl0_eng} shows that the RMS errors for models $\mathbf{HL}$, $\mathbf{AHL}$ are $0.43$ and $0.36$, respectively, which are much smaller than other predictions of SAMPL0 test set in the literature. Because of the high correlation among volume, curvatures and surface area, the utilization of model $\mathbf{AVHL}$ does not improve prediction, and its RMS error, 0.35, is slightly better than of $\mathbf{AHL}$.

			\begin{figure}[!htb]
				\begin{center}
					\includegraphics[width=0.30\columnwidth]{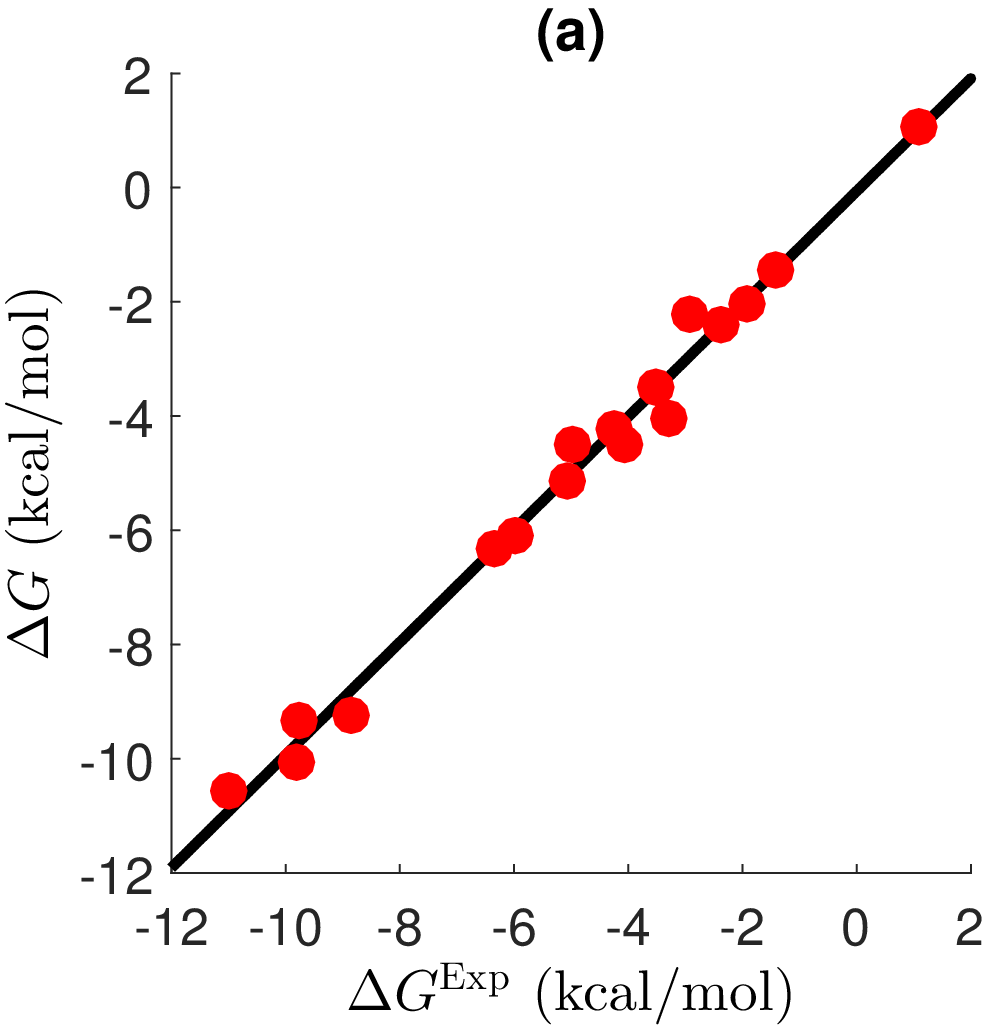}\quad
					\includegraphics[width=0.30\columnwidth]{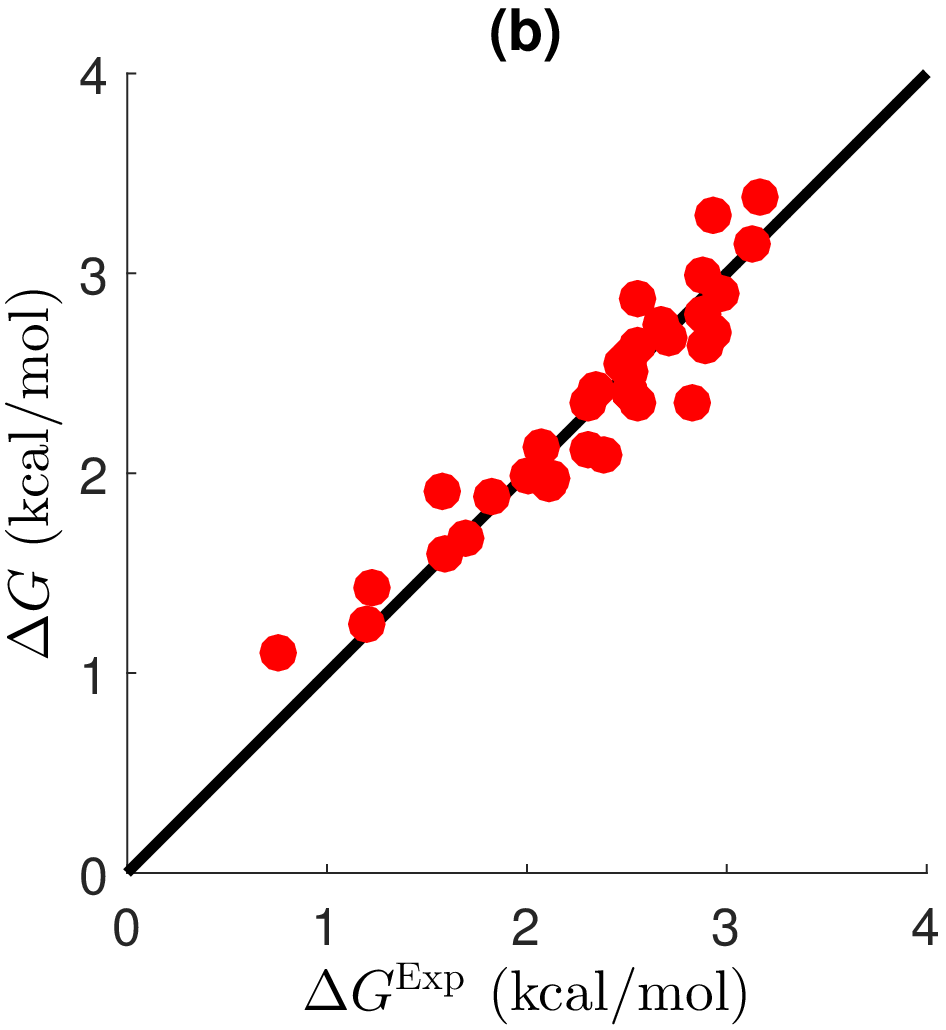}\quad
					\includegraphics[width=0.30\columnwidth]{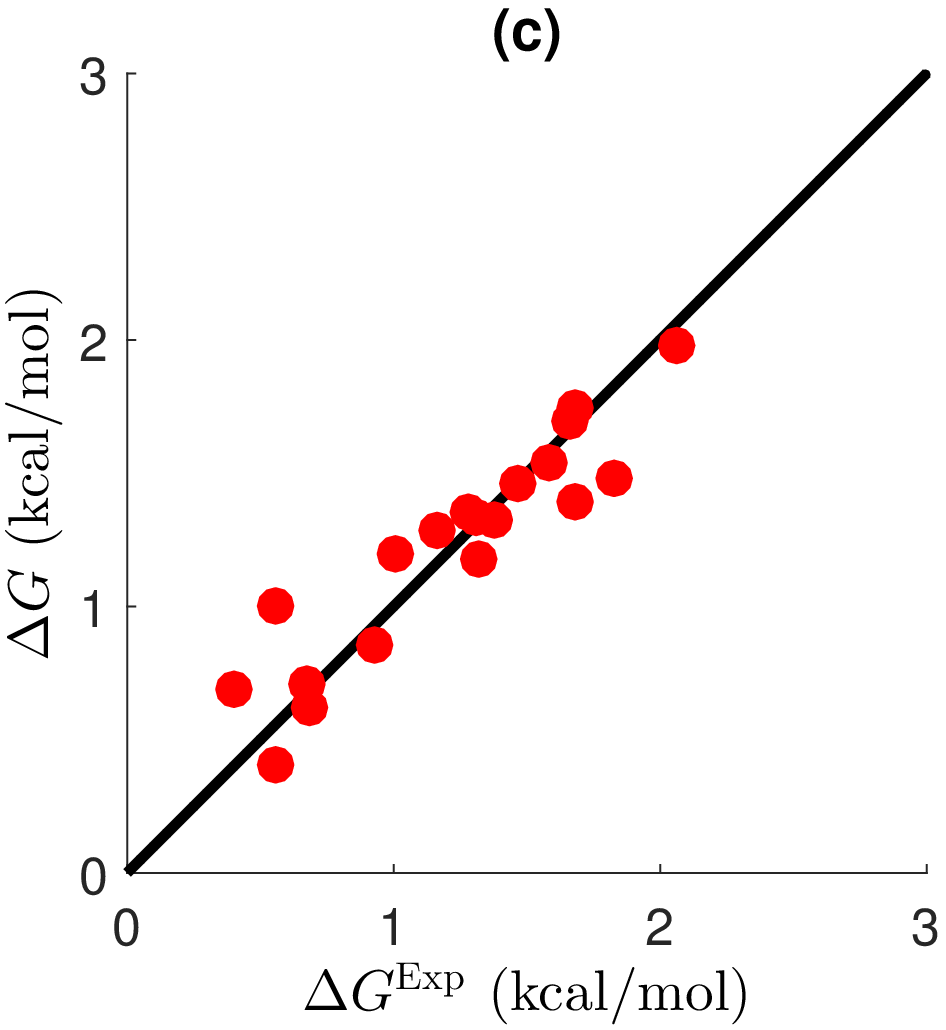}\\
					\vspace*{0.5cm}
					\includegraphics[width=0.30\columnwidth]{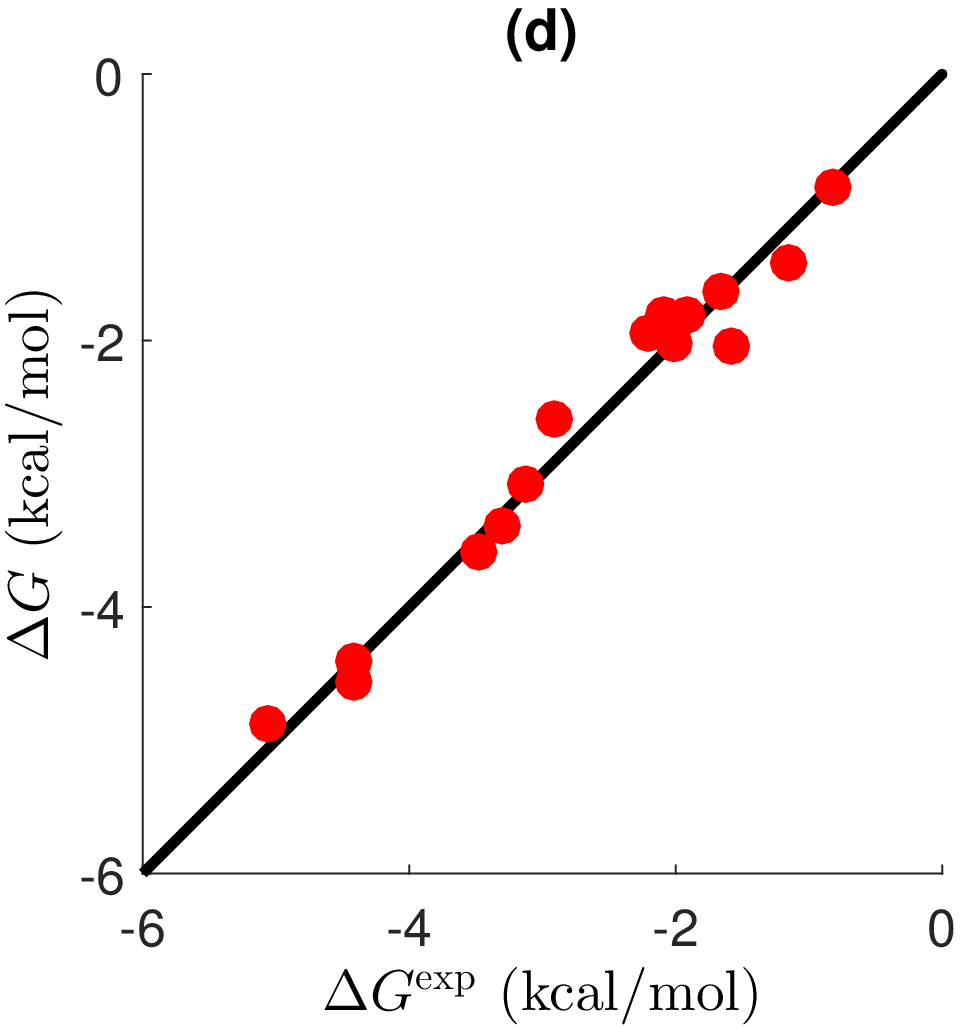}\quad
					\includegraphics[width=0.30\columnwidth]{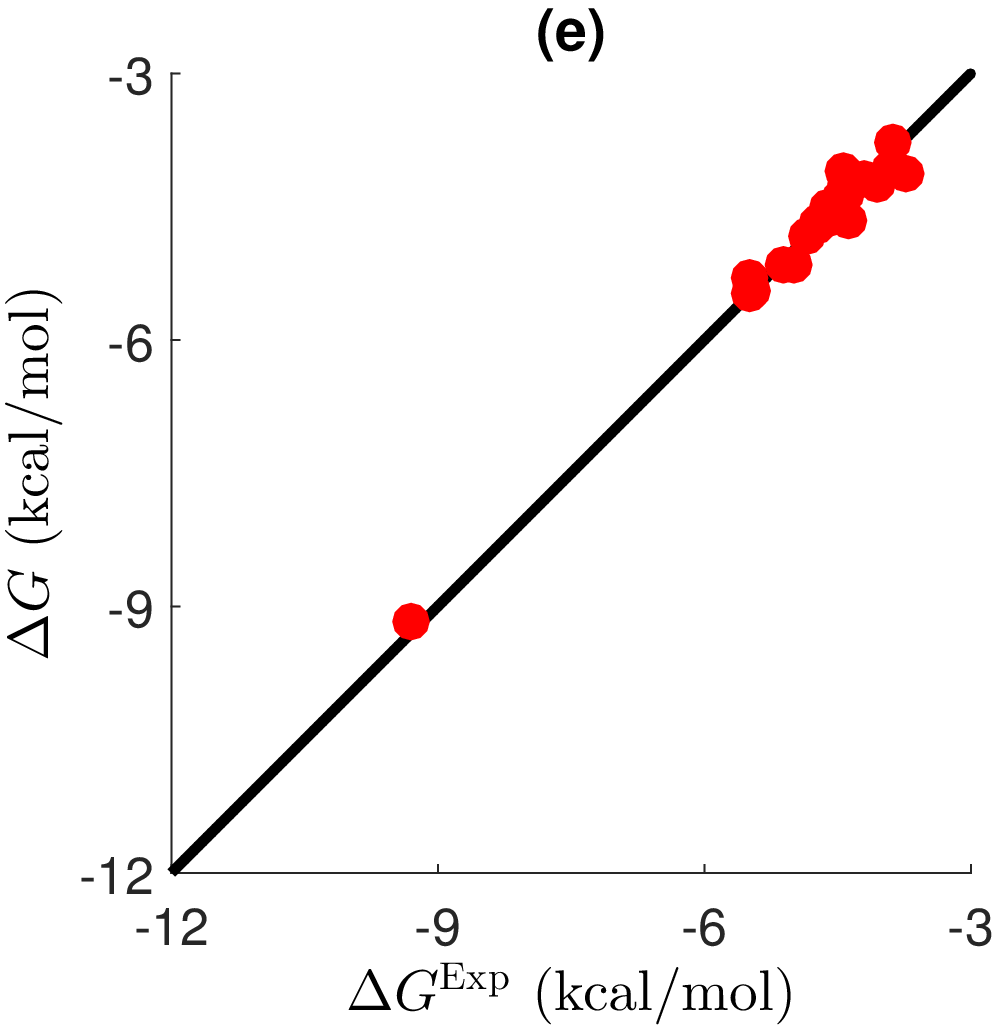}\quad
					\includegraphics[width=0.30\columnwidth]{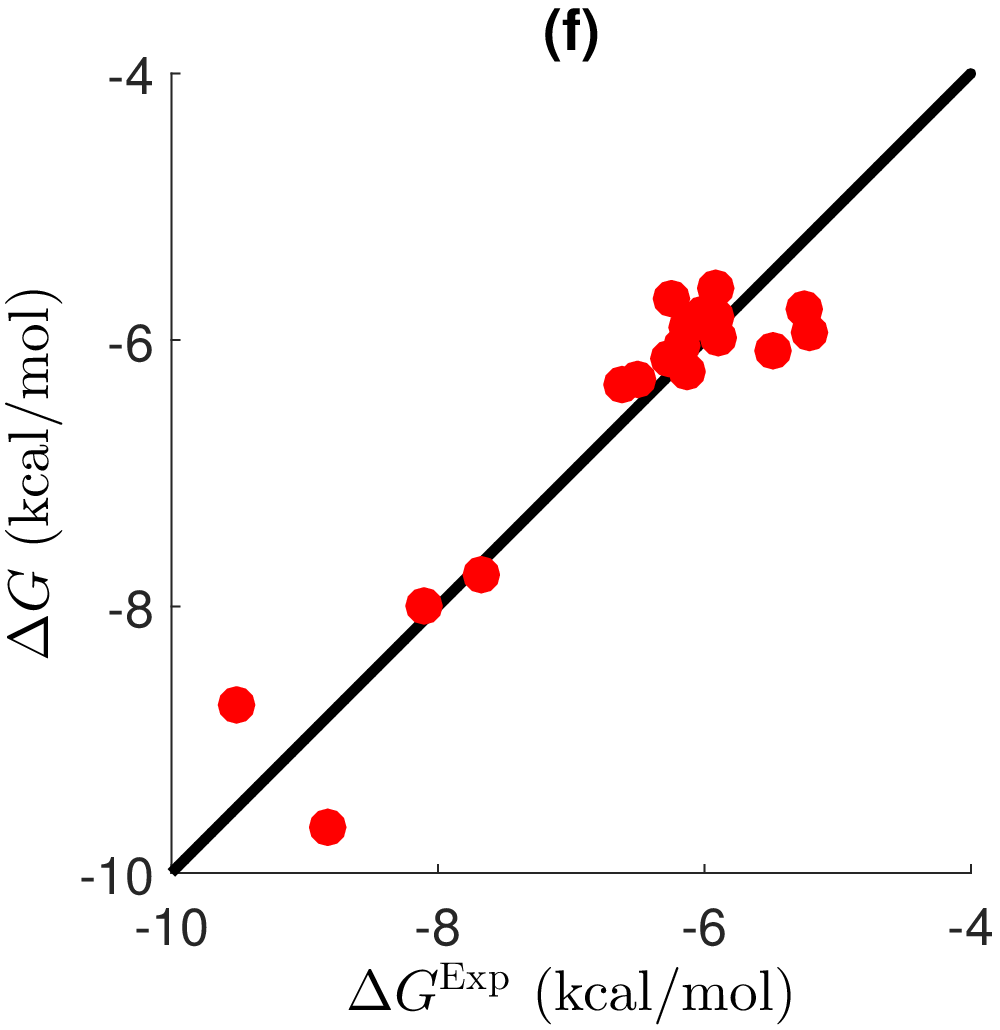}
					\caption{Comparison of $\mathbf{AVHL}$'s predicted and experiment solvation free energies for six groups. (a) SAMPL0, (b) alkene, (c) alkene, (d) ether, (e) alcohol, (f) phenol. In all charts, red circles for the predicted data, solid lines for the experiment data.}
					\label{fig.comp}
				\end{center}
			\end{figure} 
			
\subsection{The best all around model for predicting the solvation free energy}

			\begin{table}[!phtb]
				\centering
				\caption{The RMS errors (in the unit of kcal/mol) for 26 models. The highlighted numbers indicate the best RMS error in a particular category.}
				\label{tab.rms_all_models}
				\begin{tabular}{lcccccc}
					\hline 
					Model$\backslash$ Group & \multicolumn{1}{c}{SAMPL0} & \multicolumn{1}{c}{alkane} & \multicolumn{1}{c}{alkene} & 
					\multicolumn{1}{c}{ether} & \multicolumn{1}{c}{alcohol} & \multicolumn{1}{c}{phenol} \\
					\hline
					$\mathbf{A}$                &2.27               &0.40               &0.35           &0.84           &0.57           &0.59\\    
					$\mathbf{V}$                &2.34               &0.44               &0.39           &0.85           &0.62           &0.61\\    
					$\mathbf{L}$                &\textbf{1.07}      &\textbf{0.29}      &0.34           &\textbf{0.23}  &\textbf{0.28}  &\textbf{0.55}\\    
					$\mathbf{k}_1$              &2.35               &0.41               &0.33           &0.83           &0.54           &0.63\\    
					$\mathbf{k}_2$              &2.32               &0.40               &0.33           &0.81           &0.52           &0.59\\    
					$\mathbf{G}$                &2.23               &0.43               &\textbf{0.32}  &0.83           &0.54           &0.64\\    
					$\mathbf{H}$                &2.34               &0.41               &0.33        &   0.81            &0.51           &0.61\\    
					\hline
					$\mathbf{AL}$             &0.45               &0.23               &0.20  &         0.23            &0.28           &0.54\\
					$\mathbf{VL}$             &1.06               &0.28               &0.33   &        \textbf{0.19}   &\textbf{0.18}  &\textbf{0.44}\\
					$\mathbf{k}_1\mathbf{L}$  &0.66               &\textbf{0.22}      &0.19     &      0.23            &0.28           &0.48\\  
					$\mathbf{k}_2\mathbf{L}$  &0.65               &0.23               &0.23       &    0.22            &0.28           &0.54\\ 
					$\mathbf{GL}$             &0.52               &0.23               &\textbf{0.18}  &0.23           &0.28           &0.47\\  
					$\mathbf{HL}$             &\textbf{0.43}      &0.23               &0.24           &0.22           &0.28           &0.53\\  
					\hline
					$\mathbf{AVL}$                 &0.45               &\textbf{0.19}      &0.19           &0.17           &\textbf{0.17}  &0.42\\
					$\mathbf{Ak}_1\mathbf{L}$      &0.36               &0.22               &0.19           &0.22           &0.28           &0.46\\
					$\mathbf{Ak}_2\mathbf{L}$      &0.45               &0.23               &0.19           &\textbf{0.12}  &0.19           &0.53\\
					$\mathbf{AGL}$                 &\textbf{0.31}      &0.23               &0.19           &0.23           &0.27           &0.43\\
					$\mathbf{AHL}$                 &0.36               &0.22               &0.18           &0.14           &0.18           &0.53\\
					$\mathbf{Vk}_1\mathbf{L}$      &0.53               &0.21               &0.19           &0.19           &\textbf{0.17}  &\textbf{0.41}\\
					$\mathbf{Vk}_2\mathbf{L}$      &0.50               &\textbf{0.19}      &0.20           &0.18           &\textbf{0.17}  &0.42\\
					$\mathbf{VGL}$                 &0.46               &0.20               &\textbf{0.17}  &0.18           &\textbf{0.17}  &\textbf{0.41}\\
					$\mathbf{VHL}$                 &0.40               &0.20               &0.22           &0.19           &0.18           &\textbf{0.41}\\
					\hline
					$\mathbf{AVk}_1\mathbf{L}$        &0.31               &0.19               &0.18           &0.14           &0.17           &\textbf{0.41}\\
					$\mathbf{AVk}_2\mathbf{L}$        &0.45               &\textbf{0.18}      &0.19           &0.12           &0.16           &0.42\\
					$\mathbf{AVGL}$                   &\textbf{0.28}      &0.19               &\textbf{0.17}  &0.14           &0.17           &\textbf{0.41}\\
					$\mathbf{AVHL}$                   &0.35               &0.18               &0.18           &\textbf{0.11}  &\textbf{0.15}  &\textbf{0.41}\\
					\hline
					
				\end{tabular}
			\end{table}

			Finally, we    determine which model will have the best solvation free energy prediction in each group, and then which one will provide an good prediction on average. Table \ref{tab.rms_all_models} lists all the RMS errors of 26  models over 6 groups including SAMPL0, alkane, alkene, ether, alcohol and phenol sets. These results again confirm the important role of Lennard-Jones potential in the accuracy of solvation energy prediction as other studies have noted \cite{ashbaugh1999universal,Gallicchio:2000,Wagoner:2006,David:2019JCTCsolvation}. The RMS errors of model $\mathbf{L}$ for SAMPL0, alkane, alkene, ether, alcohol, and phenol sets are, respectively, $1.07$, $0.29$, $0.34$, $0.23$, $0.28$ and $0.55$. It is obvious that these predictions are still not the best performance in comparison to other work such as that in Ref. \cite{BaoWang:2015a}. This is easy to apprehend because model $\mathbf{L}$ only consists of Lennard-Jones potential while that in our previous work \cite{BaoWang:2015a} includes surface area, volume and Lennard-Jones potential itself. While models lacking of Lennard-Jones potential usually perform poorly in solvation free energy prediction. Specially, for SAMPL0 the RMS errors of those models are larger than $2.0$. However, for the rest of the test sets, the RMS errors of models without Lennard-Jones potential are  always under $0.85$. Especially, in alkene test set, model $\mathbf{G}$ delivers a better RMS error, 0.32, than that of model $\mathbf{L}$, 0.34. 
			This is probably because hydrophobic compounds in alkane and alkene groups contain only carbon and hydrogen and are very uniform. Whereas other test sets  contain oxygen or nitrogen that has  strong vdW interactions \cite{David:2019JCTCsolvation} and thus prefer the  Lennard-Jones potential.  
			
As expected, more quantities appearing in the nonpolar component will produce a better solvation prediction in general. Table \ref{tab.rms_all_models} indicates that two-term models always outperform related single-term models.  Similar patterns can be found for  three-term models  and four-term models.  The best results at each level of modeling are highlighted  in Table \ref{tab.rms_all_models}.
On average,  model $\mathbf{AVHL}$  produces the best RMS errors. Its RMS errors for six groups in the discussed order are 0.35, 0.18, 0.18, 0.11, 0.15, and 0.41, respectively. To demonstrate the accuracy of model $\mathbf{AVHL}$, Fig. \ref{fig.comp} depicts its predicted and experimental solvation free energies for SAMPL0, alkane, alkene, ether, alcohol and phenol sets. Since the results of SAMPL0 has been reported in Table \ref{tab.sampl0_eng}, in the supporting information we only list the data for alkane, alkene, ether, alcohol and phenol tests in Tables S1, S2, S3 and S4, respectively.
		
By a comparison with our earlier work \cite{ZhanChen:2010a,BaoWang:2015a},  the current models yield better solvation predictions for all test sets. 
The earlier work \cite{ZhanChen:2010a,BaoWang:2015a} employs model $\mathbf{AVL}$ and invokes sophisticated mathematical algorithms, such as differential geometry and constrained optimization. The present approach utilizes FRI based rigidity surfaces which are very simple, stable and robust. Additionally, as an intrinsic property of a protein \cite{Alvarez-Garcia:2014,Marsh:2014,Marsh:2014}, flexibility plays an important role in the solvation process. The use FRI based rigidity surfaces enables us to build the flexibility feature in our solvation analysis. 
Consequently, many of the present two-term models, such as   $\mathbf{AL}$, $\mathbf{GL}$ and $\mathbf{HL}$, are able to deliver better predictions on all test sets.  The predictions of the present $\mathbf{AVL}$ model are much better than those of our earlier $\mathbf{AVL}$ model \cite{BaoWang:2015a}. 

			
 Table \ref{tab.rms_all_models} reveals that models involving various curvatures are able to deliver some of the best results at each level of modeling.  For example, at the single-term level of modeling, the Gaussian curvature model,  $\mathbf{G}$, gives rise to better prediction for the alkene set. At the two-term level of modeling, models  $\mathbf{HL}$, $\mathbf{k}_1\mathbf{L}$ and 	$\mathbf{GL}$ provide the best predictions for SAMPL0, alkane and alkene sets, respectively. At three-term and four-term levels of modelings, most best predictions are generated by curvature based models. Since curvatures are calculated analytically in the  rigidity surface representation \cite{KLXia:2013d,Opron:2014,Opron:2015a}, the use of curvatures is very robust and simple in the present work, see Section \ref{sec.curve}.  Therefore, the present work establishes curvature  as a robust, efficient and powerful approach for solvation analysis and prediction. 

\subsection{Five-fold validation}

\begin{table}[!ht]
	\centering
	\caption{
	Training Errors (TRN. Err.) and Validation Errors (VAL. Err.) for five-fold cross validation. Errors are   in the unit of kcal/mol.}
	\label{tab.cross_val}
	\resizebox{\linewidth}{!}{%
		\begin{tabular}{ccccccccccccccc}
			\hline 
			&  \multicolumn{2}{c}{Group 1} & & \multicolumn{2}{c}{Group 2} & &\multicolumn{2}{c}{Group 3} & &\multicolumn{2}{c}{Group 4} & &\multicolumn{2}{c}{Group 5} \\
			\cline{2-3} \cline{5-6} \cline{8-9} \cline{11-12} \cline{14-15}
			& \multicolumn{1}{c}{T. Err.} & \multicolumn{1}{c}{VAL. Err.} &&  \multicolumn{1}{c}{TRN. Err.} & \multicolumn{1}{c}{VAL. Err.} &&  \multicolumn{1}{c}{TRN. Err.} & \multicolumn{1}{c}{VAL. Err.} &&  \multicolumn{1}{c}{TRN. Err.} & \multicolumn{1}{c}{VAL. Err.} &&  \multicolumn{1}{c}{TRN. Err.} & \multicolumn{1}{c}{VAL. Err.} \\
			\hline		   
			Alkane & 0.19&	0.19&&	0.17& 0.24 &&	0.18 &	0.23 &&	0.18 &	0.23&&	0.19&	0.15\\
			Alkene & 0.15&	0.40&&	0.14& 0.34 &&	0.18 &	0.30 &&	0.17 &	0.23&&	0.19&	0.10\\
			Ether  & 0.10&	0.21&&	0.11& 0.13 &&	0.10 &	0.22 &&	0.07 &	0.26&&	0.12&	0.07\\
			Alcohol& 0.15&	0.21&&	0.17& 0.07 &&	0.11 &	0.31 &&	0.14 &	0.46&&	0.14&	0.27\\
			Phenol & 0.39&	0.57&&	0.39& 0.67 &&	0.32 &	0.86 &&	0.44 &	0.32&&	0.33&	0.97\\
			\hline
		\end{tabular}	}
\end{table}
				
To further estimate how accurately the models with optimized parameters perform in practice, we carry out 5-fold cross validation. In this evaluation, each group of molecules is partitioned into 5 sub-groups as uniformly as possible. Of 5 sub-groups, we leave out one sub-group and employ model {\bf AVHL} for the rest four sub-groups of of molecules. The optimized parameters are then utilized for the left out sub-group. Table \ref{tab.cross_val} lists training errors and validation errors. It is seen that these two errors are of the same level, indicating the present method performs well.

\section{Conclusion}

Solvation analysis is a fundamental issue in computational biophysics, chemistry  and material science and has attracted much attention in the past two decades.  Implicit solvent models that split the solvation free energy into polar and nonpolar contributions have been a main workhorse in solvation free energy prediction. While the Poisson-Boltzmann theory is a well established model for   polar solvation energy prediction, there is no general consensus about what constitutes a good nonpolar component. This paper explores the impact of area, volume, curvature and Lennard-Jones potential to the accuracy of the solvation free energy prediction in conjugation with  a Poisson-Boltzmann based polar solvation model.  To this end, 26 models involving the presence of different quantities in the nonpolar component are systematically studies in the current work. Some of these models that consist of Gaussian curvature, mean curvature, minimum curvature or maximum curvature  are first known to our knowledge. 
			
In order to analytically evaluate molecular  curvatures, we utilize rigidity surfaces \cite{KLXia:2013d,Opron:2014,Opron:2015a} as the molecular surface representation.  Since the use of the rigidity surface does not require a surface evolution as in previous approaches \cite{ZhanChen:2010a,ZhanChen:2012,BaoWang:2015a}, the algorithm for achieving parameter optimization in the nonpolar component is much simpler than that in our earlier work \cite{BaoWang:2015a}. To benchmark our models, we employ the SAMPL0 test set with 17 molecules, alkane set with 35 molecules, alkene set with 19 molecules, ether set with 15 molecules, alcohol set with 23 molecules, and phenol set with 18 molecules.

We first carry out 	intensive correlation analysis. It is found  that surface areas and  surface enclosed volumes are highly correlated for the above mentioned molecules, whereas various  curvatures are poorly correlated to surface areas. Therefore, curvatures are complementary to surface areas and surface enclosed volumes in solvation modeling.    Nevertheless,  for a given set of similar molecules, maximum, minimum, mean and Gaussian  curvatures  and  Gaussian curvatures are highly correlated to each other and to surface areas.   

Based on the correlation analysis, a total 26 nontrivial models are constructed and examined against 6 test sets of molecules.  Numerous numerical experiments  indicate that   the Lennard-Jones potential is essential to the accuracy of solvation free energy prediction, especially for molecules involving strong van der Waals interactions or attractive dispersive effects. However, it is found that various curvatures are at least as useful as surface area and surface enclosed volume in nonpolar solvation modeling. Many curvature based models deliver some of the best solvation free energy predictions.


{\bf 	Supporting Information Available}

Addition results for interested models and additional correlation analysis for various curvatures (filename: URL will be inserted by publisher).
	
			
			
\begin{acknowledgement}
			This work was supported in part by NSF Grant IIS- 1302285   and  MSU Center for Mathematical Molecular Biosciences Initiative. 
\end{acknowledgement}

\bibliography{refs}

\end{document}